\documentclass[acmsmall,screen]{acmart}
\AtBeginDocument{%
  }

\newcommand{\scheme}{InterWebCoder}
\usepackage{enumitem}
\usepackage{algorithm}
\usepackage{algpseudocode}
\usepackage{booktabs}
\usepackage{tabularx}
\usepackage{array}
\usepackage[table]{xcolor}
\usepackage{pifont}

\usepackage{subfigure}
\usepackage{tcolorbox}
\newcommand{\yes}{\color{green!60!black}\ding{51}}
\newcommand{\no}{\color{red!60!black}\ding{55}} 

\setcopyright{acmlicensed}
\copyrightyear{2018}
\acmYear{2018}
\acmDOI{XXXXXXX.XXXXXXX}
\acmConference[Conference acronym 'XX]{Make sure to enter the correct
  conference title from your rights confirmation email}{June 03--05,
  2018}{Woodstock, NY}
\acmISBN{978-1-4503-XXXX-X/2018/06}

\begin{document}






\title{InterWebCoder: Learning Interactive Webpage Generation from Executable Tests via Reinforcement Learning}

\author{Jingyu Xiao}
\affiliation{%
  \institution{The Chinese University of Hong Kong}
  \city{Hong Kong}
  \country{China}
}
\email{jyxiao@link.cuhk.edu.hk}

\author{Dasen Dai}
\affiliation{%
  \institution{The Chinese University of Hong Kong}
  \city{Hong Kong}
  \country{China}
}
\email{daidasen@link.cuhk.edu.hk}

\author{Yuxuan Wan}
\affiliation{%
  \institution{The Chinese University of Hong Kong}
  \city{Hong Kong}
  \country{China}
}
\email{yxwan9@cse.cuhk.edu.hk}

\author{Jen-tse Huang}
\affiliation{%
  \institution{Johns Hopkins University}
  \city{Baltimore}
  \country{USA}
}
\email{jhuan236@jh.edu}

\author{Yintong Huo}
\authornote{Yintong Huo is the corresponding author.}
\affiliation{%
 \institution{Singapore Management University}
 \city{Singapore}
 \country{Singapore}}
\email{ythuo@smu.edu.sg}

\author{Hongyu Zhang}
\affiliation{%
  \institution{Chongqing University}
  \city{Chongqing}
  \country{China}}
\email{hongyujohn@gmail.com}

\author{Michael R. Lyu}
\affiliation{%
  \institution{The Chinese University of Hong Kong}
  \city{Hong Kong}
  \country{China}}
\email{lyu@cse.cuhk.edu.hk}

\renewcommand{\shortauthors}{Jingyu Xiao et al.}

\begin{abstract}
Interactive webpages underpin modern applications, but reproducing both their visual appearance and multi-step behavior remains challenging for multimodal large language models (MLLMs). Scalable training data must capture user actions, state transitions, and expected outcomes, while real-world webpage code is often entangled with irrelevant scripts and external dependencies. Moreover, existing GUI-agent-based verification provides slow and potentially unstable functional feedback.
To address these challenges, we propose \scheme{}, an efficient framework for interactive webpage code generation. \scheme{} introduces a Test-Specification-Guided and Execution-Verified Data Synthesis pipeline that records real-world interaction trajectories and synthesizes self-contained webpage code with executable test scripts. A failure-attributed repair mechanism uses runtime evidence to locate and iteratively fix errors in either the webpage code or test script. We then combine test-ID-weighted supervised fine-tuning with group-relative policy optimization, directly optimizing functional correctness and visual fidelity using test pass rates and state-level visual similarity. This replaces costly GUI-agent verification with fast, reproducible executable feedback. Our pipeline produces 1,736 verified samples from 2,058 tasks, achieving an 84.35\% synthesis success rate. On IWR-Bench and Vision2Web, \scheme{} achieves final scores of 48.54 and 48.42 with a 9B backbone, outperforming 30B-level models and Gemini-3.7-Flash while remaining competitive with Claude-Sonnet-5. It reduces reward-computation latency by over 97\%, eliminates external MLLM verification costs, and accelerates end-to-end training by up to \(1.4\times\).

\end{abstract}




\maketitle

\section{Introduction}

Websites have become an indispensable interface through which people access information, communicate with others, and interact with digital services. However, developing a high-quality website remains a time-consuming and labor-intensive process, requiring developers to translate design specifications into executable code while carefully maintaining visual fidelity~\cite{GUIPilot}, functional correctness~\cite{xiao2024interaction2code}, and cross-component consistency.

Multimodal Large Language Models (MLLMs) have demonstrated strong
vision-language understanding capabilities, driving growing interest in UI-to-code generation. Representative methods include
DCGen~\cite{wan2024automatically} for modular decomposition,
DeclarUI~\cite{zhou2024bridging} for segmentation and transition modeling, UICopilot~\cite{gui2025UICopilot} for hierarchical generation, and LatCoder~\cite{gui2025latcoder} for layout-aware reconstruction. While these methods primarily reconstruct individual UI states, real-world web applications involve interactions that connect multiple states. Interactive webpage generation~\cite{xiao2024interaction2code, chen2026iwr} therefore aims to translate a task instruction and a sequence of visual interaction states into executable HTML, CSS, and JavaScript. Each state captures the webpage before or after a user action, such as clicking a button, entering text, or selecting an option. The generated code must reproduce not only the appearance of each state but also the interaction logic governing the transitions between them.

However, learning these visual and behavioral
requirements jointly, introduces substantial challenges in data
construction and model training.






\textbf{First, constructing large-scale, high-quality training data is difficult.} Unlike static UI2Code data, interactive examples must capture user actions, state transitions, and expected outcomes. On the interaction side, constructing multi-state prototypes requires costly human annotation of user actions, state transitions, and expected outcomes~\cite{xiao2024interaction2code,
chen2026iwr}. On the code side, real-world webpages are entangled with irrelevant scripts and external dependencies~\cite{gui2025webcode2m}, making it difficult to obtain clean, self-contained training targets.


\textbf{Second, obtaining reliable functional feedback is costly and time-consuming.} MLLMs have demonstrated the ability to perform GUI testing~\cite{teoh2026webtestpilot}. Existing interactive webpage generation methods such as TDDev~\cite{wan2026runnable} and WebGen-Agent~\cite{lu2026webgen} similarly employ GUI agents to execute and verify multi-step interactions. However, this process requires repeated, time-consuming MLLM calls and remains vulnerable to errors in visual grounding, action execution, and state interpretation.

We provide a detailed empirical analysis of these two
challenges in Section~\ref{sec:motivation}. To address these challenges, we propose \scheme, an efficient data synthesis and training framework for interactive webpage generation.


First, we introduce a \textbf{Test-Specification-Guided and
Execution-Verified Data Synthesis (TEDS) pipeline}. TEDS uses a GUI agent to interact with real websites and automatically records the actions and resulting visual states. The recorded actions define the required operations, while the observed state changes provide the expected outcomes for constructing the test specification. Because real-world source code is often inaccessible, noisy, and dependent on external services, a coding agent
reconstructs a clean, self-contained webpage and generates a corresponding executable test script. TEDS executes the script on the synthesized webpage and verifies the result by checking both its assertions and the visual similarity between the recorded and replayed states. When verification fails, a failure-attributed repair mechanism determines whether the error originates from the webpage code or the test script and repairs the identified source of failure.This
pipeline constructs verified training samples without manually implementing webpages, writing tests, or labeling execution outcomes.

Second, we propose a \textbf{Test-Driven Reinforcement Learning (TDRL) framework} that uses executable test scripts to provide deterministic rewards, replacing costly and unstable GUI-agent verification.  Supervised fine-tuning primarily learns token-level code patterns and does not explicitly determine whether the generated interactions execute correctly. TDRL executes the tests generated by TEDS on each webpage and provides reproducible rewards based on the proportion of passed test steps and the visual similarity between the rendered and reference states.




Consequently, TEDS addresses the shortage of clean, behaviorally verified training data, while TDRL provides an efficient mechanism for learning from such data and directly optimizing functional correctness and visual fidelity. Our contributions are summarized as follows:



\begin{itemize}[leftmargin=*, nosep]

    \item We introduce a Test-Specification-Guided and Execution-Verified Data Synthesis (TEDS) pipeline that synthesizes clean webpage code and executable test scripts from real-world interaction trajectories. TEDS attributes execution failures to either the webpage code or test script using rule-based diagnostics, and then iteratively repairs the responsible component. This process enables scalable construction of verified training data without manual annotation.
    

    \item We propose Test-Driven Reinforcement Learning (TDRL), which replaces costly and unstable GUI-agent verification with fast, reproducible execution-based rewards, enabling direct and efficient optimization of interactive behavior and visual fidelity.
    

    \item We conduct extensive experiments on IWR-Bench~\cite{chen2026iwr} and Vision2Web~\cite{he2026visionweb} with MLLMs of different scales, demonstrating that our method improves both functional correctness and visual fidelity while achieving substantial gains in training efficiency.
\end{itemize}

\section{Background and Motivation}

\subsection{Interactive Prototyping in Web Development}

Web development generally follows a design-implementation workflow. Designers use tools such as Figma~\cite{figma}, Sketch~\cite{sketch}, and Axure RP~\cite{axure} to create interactive prototypes, which developers translate into HTML, CSS, and JavaScript. Unlike static mockups, these prototypes connect interface states through actions such as clicking, typing, and navigation, specifying both visual appearance and expected behavior. Formally, an interactive prototype is represented as $\mathcal{S}=\{I_0 \xrightarrow{a_1} I_1 \xrightarrow{a_2}\cdots\xrightarrow{a_T} I_T\}$, where \(I_t\) is the webpage state after action \(a_t\). 


\subsection{Interactive Webpage Code Generation Task Definition}

The interactive webpage code generation task involves translating a sequence of visual interaction states into executable HTML, CSS, and JavaScript code that reproduces both the appearance and behavior of the reference webpage. Let \(\mathcal{S}=\{I_0,I_1,\ldots,I_T\}\) denote the input interactive prototype, where \(I_0\) is the initial webpage state and \(I_t\) represents the state after the \(t\)-th interaction. Given \(\mathcal{S}\), a Multimodal Large Language Model \(M\) generates webpage code $W_g=M(\mathcal{S})$. Executing the reference interaction sequence on \(W_g\) produces a replayed state sequence \(\hat{\mathcal{S}}=\{\hat{I}_0,\hat{I}_1,\ldots,\hat{I}_T\}\). The generated webpage should correctly implement the expected interactions, while each replayed state \(\hat{I}_t\) should closely approximate its reference state \(I_t\) in visual appearance.



\subsection{Motivating Challenges}
\label{sec:motivation}

We further analyze two practical bottlenecks that hinder scalable model training for interactive webpage generation: obtaining clean,
interaction-aware supervision and efficiently verifying.

\begin{figure*}[h]
\centering
\includegraphics[width=.95\textwidth]
{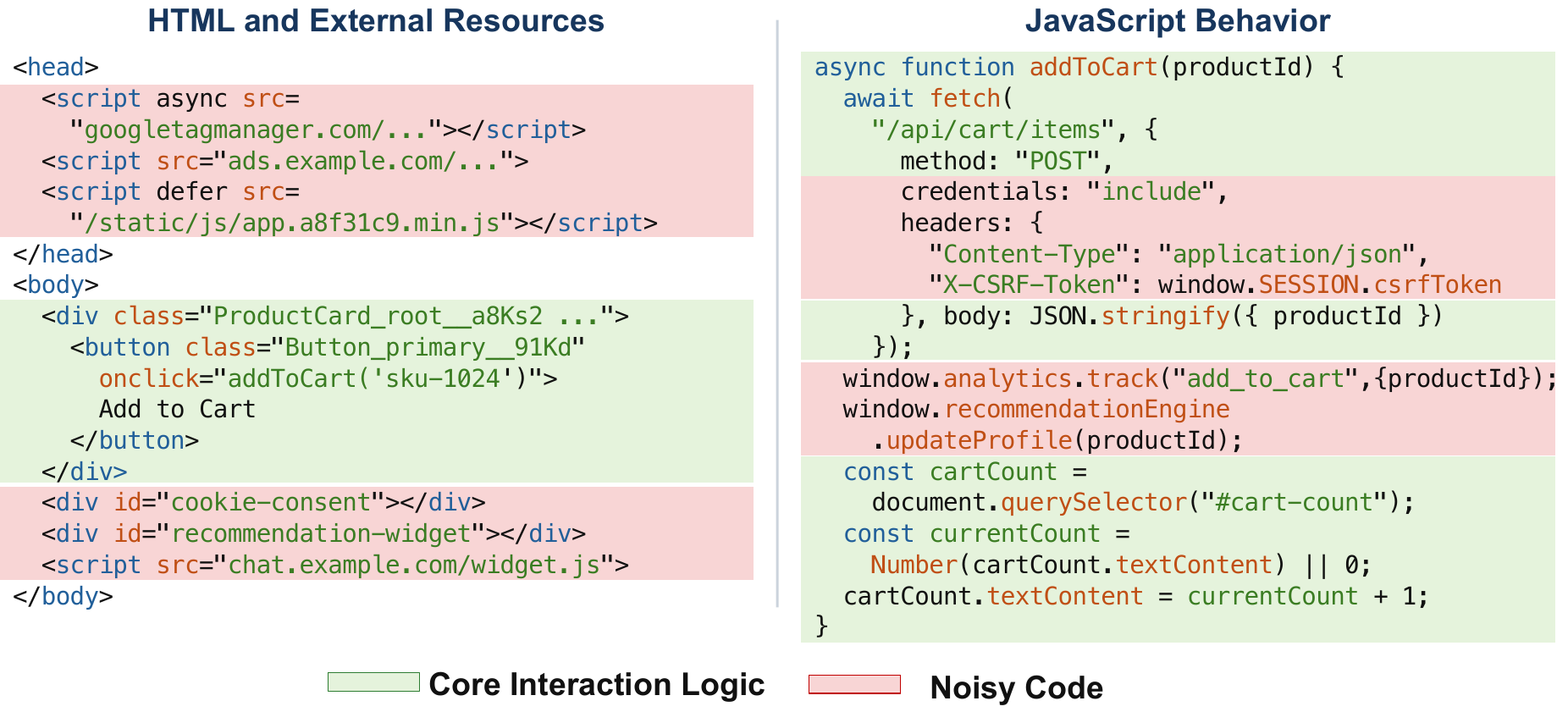}
\caption{An example of noisy real-world webpage code, where the target
add-to-cart behavior is entangled with unstable implementation details,
external dependencies, and unrelated third-party services.}
\label{fig:noisy-web-code}
\end{figure*}

\textbf{Noisy and difficult-to-scale training supervision.}
Interactive examples must capture webpage appearance, actions, state
transitions, and expected outcomes. Existing datasets rely heavily on costly manual annotation~\cite{xiao2024interaction2code,chen2026iwr}, while directly reusing real-world source code introduces substantial noise. As shown in Fig.~\ref{fig:noisy-web-code}, even a simple add-to-cart interaction can be entangled with tracking scripts, dynamic DOM attributes, session state, external APIs, and unrelated services. Since only a small portion implements the target behavior, isolating clean and executable supervision is difficult. This motivates synthesizing self-contained webpages from recorded behavior.


\textbf{Expensive and unstable functional feedback.}
Interactive correctness cannot be determined from code compilation or a single rendered screenshot. Existing methods therefore employ GUI agents to execute actions and inspect the resulting webpage states. However, this process introduces substantial latency and may fail because of incorrect visual grounding and action execution.

To quantify this overhead, we sample 102 websites from
InfiniteWeb~\cite{zhang2026infiniteweb} and conduct reinforcement learning for interactive webpage generation. We use GPT-5.6-Luna
~\cite{openai2026gpt56luna}, equipped with the Browser Use
scaffold~\cite{browser_use2024}, as the GUI-agent reward verifier. We record the time spent on rollout generation, GUI-agent reward computation, and policy updates, together with the number of interaction steps and the GUI-agent success rate.

\begin{figure*}[h]
\centering
\subfigure[Time distribution.]{
    \includegraphics[width=.19\textwidth]
    {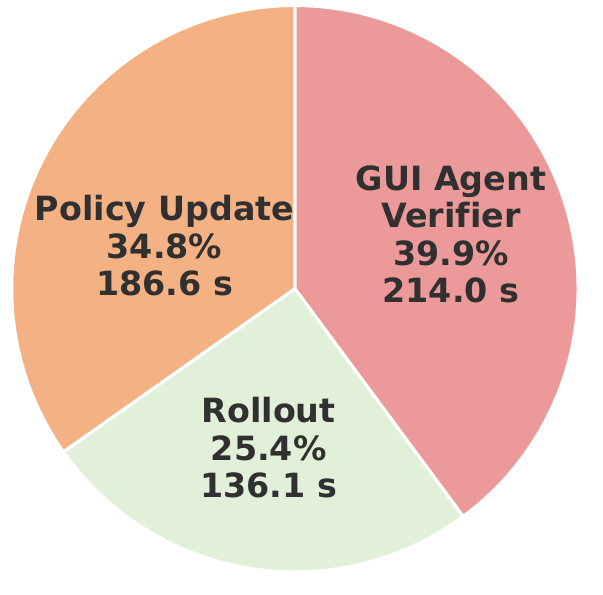}
}
\subfigure[Interaction steps.]{
    \includegraphics[width=.24\textwidth]
    {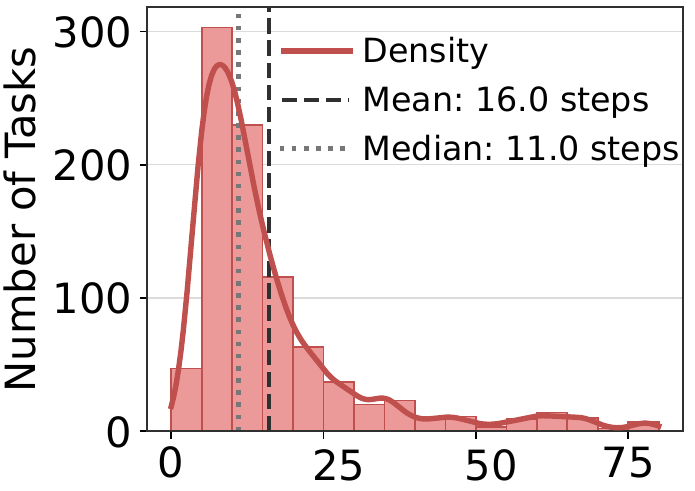}
}
\subfigure[Average duration.]{
    \includegraphics[width=.24\textwidth]
    {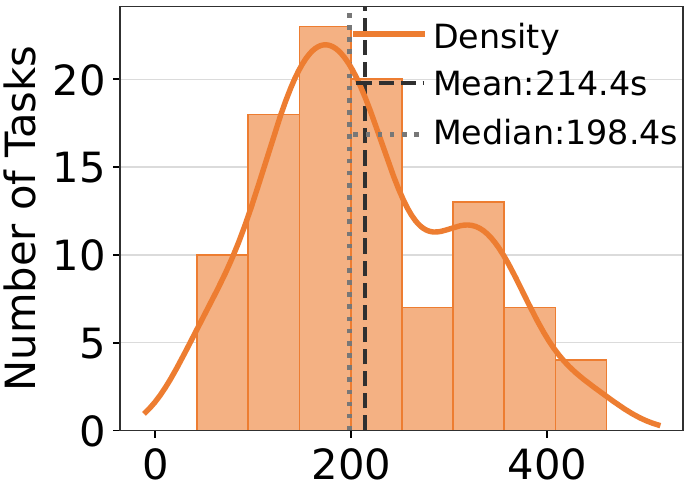}
}
\subfigure[Success rate.]{
    \includegraphics[width=.24\textwidth]
    {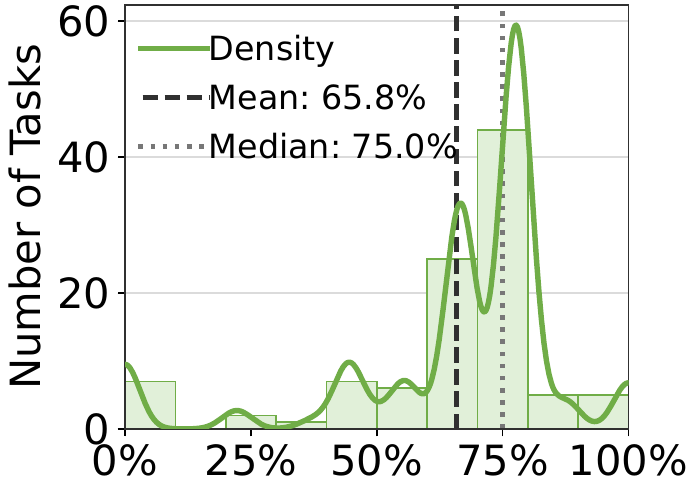}
}
\caption{Runtime and reliability of GUI-agent-based RL verification.}
\label{fig:gui-agent-overhead}
\end{figure*}

As shown in Fig.~\ref{fig:gui-agent-overhead}, GUI-agent reward computation accounts for the largest proportion of training time (39.9\%) and requires 214.0 seconds per sample on average. Each task contains 16 interaction steps on average, while the mean GUI-agent success rate is only 65.8\%. These results show that GUI-agent verification is both costly and unreliable, motivating the use of executable test scripts as a faster and more reproducible source of functional feedback.

\section{Overview}

To address the difficulty of constructing clean, scalable training data and the high cost of GUI-agent-based training feedback, our framework consists of two stages: test-specification-guided and execution-verified data synthesis, followed by test-driven model training.

\begin{figure}[ht]
\centering
\includegraphics[width = .95\textwidth]{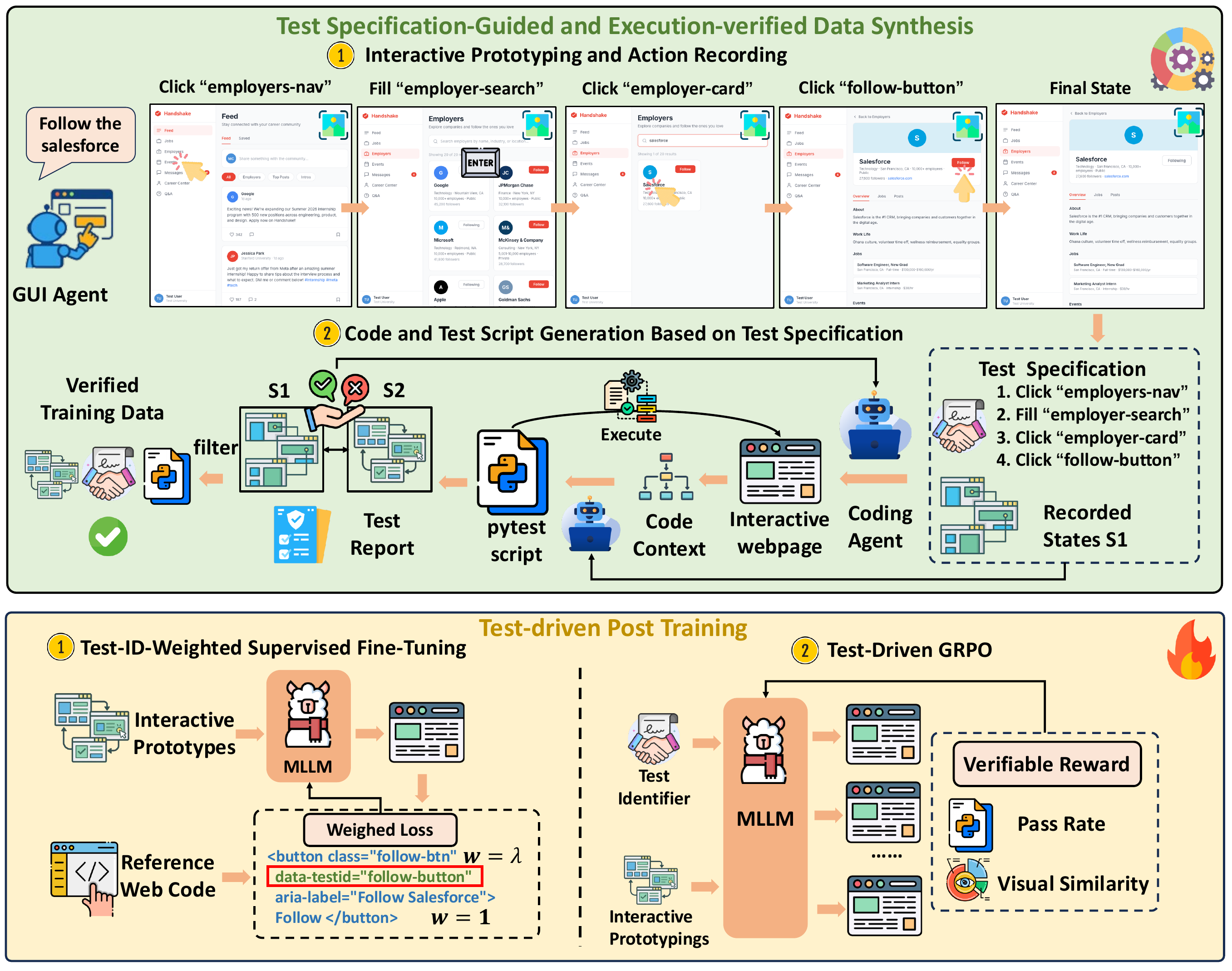}
\caption{Overview of \scheme.}
\label{fig:overview}
\end{figure}

First, a GUI agent interacts with real websites and records actions, states \(S_1\), and test specifications. A coding agent then synthesizes a clean, self-contained webpage, while the specification and relevant code context guide executable Pytest generation. Executing the script produces replayed states \(S_2\), and each sample is verified using test results and the visual consistency between \(S_1\) and \(S_2\). Failure-attributed repair corrects errors in the webpage or test script, and only verified samples are retained. Second, we reduce the cost and instability of GUI-agent-based feedback through test-driven model training, comprising Test-ID-Weighted SFT and GRPO. During SFT, we assign higher loss weights to tokens that form the element
identifiers referenced by executable tests, such as the value in \texttt{data-testid="follow-button"}.
This encourages the model to preserve reliable bindings between generated interface elements and test actions. During GRPO, the MLLM takes only the interactive prototypes as input and samples multiple candidate webpages. The test specification and executable script are used by the verifier rather than provided to the MLLM. Each candidate receives verifiable rewards based on test pass rate and visual similarity. This design replaces costly GUI-agent rollouts with fast and reproducible executable feedback.

\section{Test-Specification-Guided and Execution-Verified Data Synthesis}

\subsection{Data Collection}

\subsubsection{Website and task curation.}
We recruited four experienced PhD students to curate a diverse collection of popular websites. The annotators first developed a taxonomy covering common web domains and then independently searched for widely used websites within each category. Their selections were consolidated through discussion to remove duplicates and ensure domain diversity. Representative websites include Gmail for communication, AliExpress for retail, Coursera for education, Skyscanner for travel, and PayPal for finance.


For each selected website, we prompt an LLM with its description and visible webpage content to synthesize between one and ten user-oriented task instructions covering its common interactive functionalities. Each instruction specifies an observable user goal rather than a predefined action sequence, allowing different interaction trajectories to accomplish the same task. We automatically filter duplicated, infeasible, and security-sensitive instructions. This process requires no manual annotation of webpage code, element selectors, action sequences, or executable tests.


\subsubsection{Data statistics.}
\begin{figure}[h]
    \subfigure[Task Distribution.]{
    \label{subfig:task_distribution}
    \centering
    \includegraphics[width = .3\textwidth]{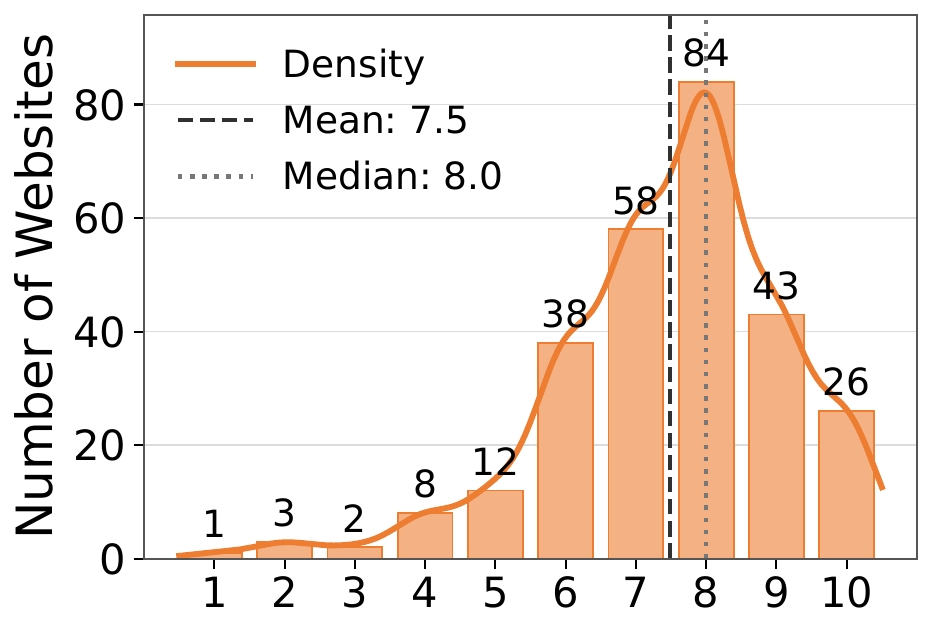}
    }
    \subfigure[Interaction Step Distribution.]{
    \label{subfig:interaction_distribution}
    \centering
    \includegraphics[width = .3\textwidth]{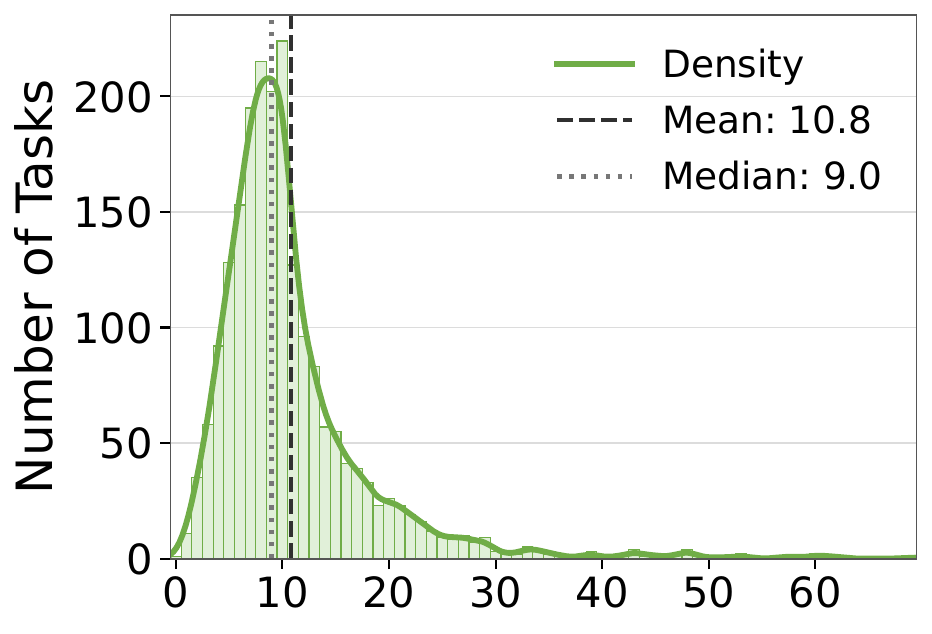}
    }
    \subfigure[Action Type.]{
    \label{subfig:action_distribution}
    \centering
    \includegraphics[width = .32\textwidth]{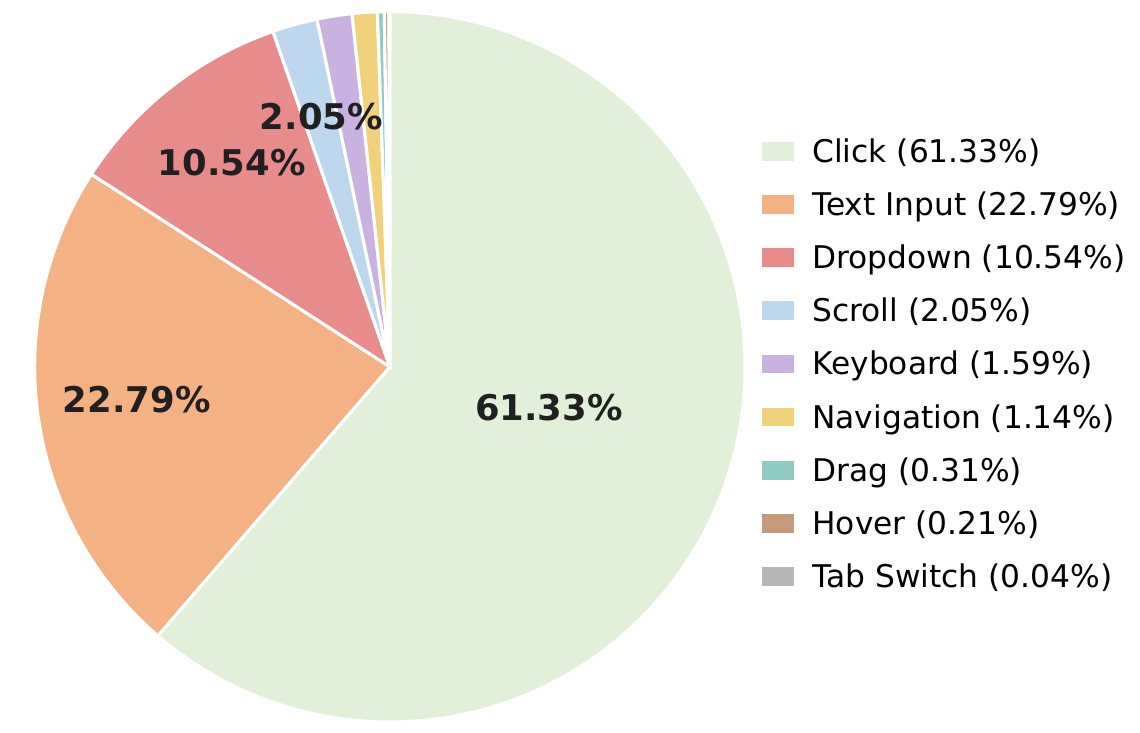}
    }
    \caption{Statistics of the collected interaction tasks.}
    \label{fig:data_distribution}
\end{figure}

The resulting collection contains 2,058 tasks from 275 websites. Table~\ref{tab:domain_distribution} reports the distribution across 15 primary domains. Community accounts for the largest proportion (15.74\%), followed by Media (10.64\%) and Technology (10.59\%), while the remaining tasks span business, health, retail, education, finance, travel, government, and other practical domains. As shown in Figure~\ref{subfig:task_distribution}, each website contains 7.5 tasks on average, with a median of 8. 




\begin{table}[h]
\centering
\caption{Distribution of primary task domains.}
\label{tab:domain_distribution}
\begin{tabular}{l r l r l r}
\toprule
\textbf{Domain} & \textbf{Proportion}
& \textbf{Domain} & \textbf{Proportion}
& \textbf{Domain} & \textbf{Proportion} \\
\midrule
Community    & 15.74\% & Retail       & 6.17\% & Real Estate & 5.00\% \\
Media        & 10.64\% & Education    & 5.88\% & Science     & 4.86\% \\
Technology   & 10.59\% & Industry     & 5.83\% & Travel      & 4.71\% \\
Business     &  8.16\% & Finance      & 5.78\% & Careers     & 2.72\% \\
Health       &  6.51\% & Productivity & 5.15\% & Government  & 2.24\% \\
\bottomrule
\end{tabular}
\end{table}

\subsection{Interactive Prototyping and Action Recording}

The first stage of our data synthesis pipeline constructs interaction prototypes from real-world webpages. Given a natural-language task instruction \(x\), a GUI agent interacts with a target webpage and completes the specified task through a sequence of actions. At each step, we record the action and capture the resulting webpage state, forming an interaction trajectory:
\begin{equation}
    \mathcal{T}
    =
    \left\{
    \left(s_t, a_t\right)
    \right\}_{t=1}^{T},
    \qquad
    \mathcal{S}^{(1)}
    =
    \left\{
    s_0, s_1, \ldots, s_T
    \right\},
\end{equation}
where \(s_0\) is the initial state, \(a_t\) denotes the \(t\)-th interaction, and \(s_t\) is the screenshot after executing \(a_t\). We refer to \(\mathcal{S}^{(1)}\) as the recorded state sequence. In the example of Fig.~\ref{fig:overview}, the agent follows Salesforce on a professional networking platform by clicking the \texttt{employers-nav} entry, entering ``salesforce'' into the
\texttt{employer-search} field, selecting the corresponding
\texttt{employer-card}, and clicking the \texttt{follow-button}. The final state confirms that Salesforce has been successfully followed.



Figures~\ref{subfig:interaction_distribution} and
\ref{subfig:action_distribution} summarize the recorded interaction trajectories. Each task contains 10.8 interaction steps on average, with a median of 9, while the long-tailed distribution includes complex trajectories with more than 60 steps. Click is the most frequent action type (61.33\%), followed by text input (22.79\%) and dropdown selection (10.54\%). The remaining actions include scrolling, keyboard operations, navigation, dragging, hovering, and tab switching. These statistics demonstrate that the collected prototypes cover both short, common interactions and long, compositional workflows.

\subsection{Webpage Code and Test Script Generation Based on Test Specification}

To represent the behavioral requirements of each trajectory, we convert the recorded actions and state changes into an executable test specifications. Each step in the test specification is defined as $ c_t =
    \left(
    o_t,\,
    z_t,\,
    v_t,\,
    \mathcal{A}_t
    \right),$ where \(o_t\) is the action type, such as \texttt{click} or \texttt{fill}, \(z_t\) is a canonical identifier assigned to the target element, \(v_t\) is an optional input value, and \(\mathcal{A}_t\) contains the expected observable changes after the action. The complete test specification is denoted by
\(\mathcal{C}=\{c_t\}_{t=1}^{T}\).


\begin{figure}[h]
\centering
\includegraphics[width = .98\textwidth]{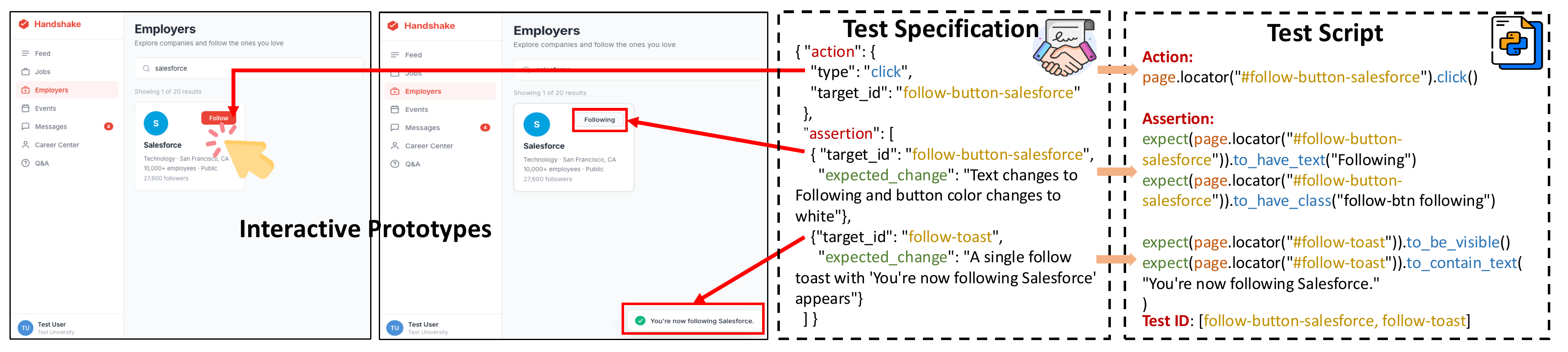}
\caption{Examples of interactive prototype, test specification and test script.}
\label{fig:test-id-example}
\end{figure}


The test specification defines both the interaction and its expected outcome. As shown in Fig.~\ref{fig:test-id-example}, clicking \texttt{follow-button-salesforce} should change ``Follow'' to ``Following'' and display a confirmation toast. By retaining only task-relevant actions and observable outcomes, it provides compact, implementation-independent guidance for webpage and test-script generation.


Given the recorded state sequence \(\mathcal{S}^{(1)}\), and its test specification \(\mathcal{C}\), a coding agent synthesizes a clean and self-contained interactive webpage:
$
    \hat{W}
    =
    f_{\mathrm{code}}
    \left(
    x,\mathcal{S}^{(1)},\mathcal{C}
    \right).
$
The generated implementation is required to reproduce the visual states in \(\mathcal{S}^{(1)}\) while satisfying all interactions specified by \(\mathcal{C}\). To establish a reliable correspondence between the specification and the generated code, every referenced component is assigned its canonical target identifier. External dependencies, tracking scripts, advertisements, and other task-irrelevant content are excluded, resulting in clean and executable training code.

\textbf{Code Context-enhanced Test Script Generation}.
Generating a reliable test script directly from screenshots is difficult because visual observations do not expose concrete selectors or implementation details. We therefore introduce context-enhanced test-script generation. Specifically, we extract the code context \(\mathcal{G}(\hat{W})\), including the relevant component hierarchy, target identifiers, and interaction handlers, and generate a Pytest script conditioned on both the test specification and the synthesized implementation:
$
    \hat{P}
    =
    f_{\mathrm{test}}
    \left(
    \mathcal{C},
    \mathcal{G}(\hat{W})
    \right).
$
For each specification step, the generated script first locates the target element and executes the specified action, and then translates the expected changes into explicit assertions. For instance, the specification for following a company is instantiated as a click on \texttt{\#follow-button-salesforce}, followed by assertions that the button displays ``Following'' and that the confirmation toast becomes visible with the expected message. Conditioning on the code context prevents invalid selectors and ensures that the test script remains aligned with the generated webpage.


\subsection{Failure-Attributed Repair}

\begin{table*}[t]
\centering
\caption{Failure taxonomy and rule-based attribution.}
\label{tab:triage-failure-types}
\small
\setlength{\tabcolsep}{5pt}
\renewcommand{\arraystretch}{0.9}

\begin{tabularx}{\textwidth}{
    @{}
    >{\raggedright\arraybackslash\ttfamily\footnotesize}p{3.8cm}
    >{\raggedright\arraybackslash}p{2.2cm}
    >{\raggedright\arraybackslash}X
    @{}
}
\toprule
\multicolumn{1}{@{}l}{\textbf{Failure Type}} &
\textbf{Attributed Component} &
\textbf{Evidence} \\
\midrule

\rowcolor{gray!12}
\multicolumn{3}{@{}c}{\textbf{Static Time}} \\
\addlinespace[2pt]

TARGET\_DEFINITION\_ERROR &
\textrm{Webpage Code} &
A specification-required target ID is missing from the Webpage Code or assigned
to multiple elements. \\

\addlinespace[2pt]
TARGET\_REFERENCE\_ERROR &
\textrm{Test Script} &
A specification-required target is not referenced by the Test Script, or the Test
Script references an out-of-specification target. \\

\addlinespace[2pt]
SCRIPT\_DOM\_MISMATCH &
\textrm{Test Script} &
An action or assertion is incompatible with the source-level DOM element,
such as applying \texttt{select} to a non-\texttt{select} element. \\

\addlinespace[2pt]
SCRIPT\_SPECIFICATION\_MISMATCH &
\textrm{Test Script} &
The action order, action target, or assertion target is inconsistent with the
interaction specification. \\

\addlinespace[3pt]
\midrule
\rowcolor{gray!12}
\multicolumn{3}{@{}c}{\textbf{Runtime}} \\
\addlinespace[2pt]

TARGET\_AVAILABILITY\_ERROR &
\textrm{Webpage Code} &
A specification-required target is unexpectedly absent, duplicated, hidden,
disabled, or otherwise non-interactable after the validated action prefix
successfully executes. \\


\addlinespace[2pt]
SCRIPT\_RUNTIME\_ERROR &
\textrm{Test Script} &
The Test Script fails during syntax checking, importing, test collection, or
initialization before trajectory execution. \\

\addlinespace[2pt]
BEHAVIORAL\_MISMATCH &
\textrm{Uncertain} &
An action or assertion failure cannot be deterministically attributed to
Webpage Code behavior or Test Script semantics or timing. \\

\bottomrule
\end{tabularx}
\end{table*}


Although the webpage code \(\hat{W}\) and test script \(\hat{P}\) are generated from the same test specification, they may still be mutually inconsistent or fail during execution. A failure can originate from an incorrect webpage implementation, an invalid test script, or an ambiguous interaction between the two. Regenerating both artifacts whenever a failure occurs is inefficient and may corrupt the component that is already correct. We therefore introduce a failure-attributed repair procedure that identifies the responsible component before performing targeted correction.

\textbf{Rule-based failure attribution.}
We establish the failure taxonomy in Table~\ref{tab:triage-failure-types}, covering failures detected during static analysis and browser execution. Before executing the webpage, we compare the test specifications, webpage source, and test script to identify deterministic inconsistencies. Specifically, we verify that every specification-required target ID is defined exactly once in the webpage code and is correctly referenced by the test script. We also examine the action order, action targets, assertion targets, and compatibility between test APIs and source-level DOM elements. Missing or duplicated target definitions are attributed to the webpage code, whereas invalid target references, specification inconsistencies, and incompatible test operations are attributed to the test script. If the pair passes static analysis, we execute the test script in a browser environment. For each failure, we record the successfully executed action prefix, the first failing action or assertion, the runtime DOM state, and the corresponding exception. If a required target becomes absent, duplicated, hidden, disabled, or otherwise non-interactable after the preceding actions succeed, the failure is attributed to the webpage code. Syntax errors, import failures, test-collection failures, initialization errors, and invalid runtime API bindings are attributed to the test script.


\textbf{LLM-based attribution for uncertain failures.}
Failures that cannot be deterministically attributed, such as an assertion timeout caused by either an incorrect webpage transition or an overly strict test condition, are categorized as \texttt{BEHAVIORAL\_MISMATCH}. For these cases, an LLM judge analyzes the specification, generated artifacts, execution trace, runtime DOM, and screenshots to identify the faulty component and provide repair guidance.

\textbf{Targeted iterative repair.}
We repair only the attributed component: webpage failures trigger modifications to the relevant HTML, CSS, or JavaScript, whereas test-script failures trigger corrections to selectors, actions, assertions, or synchronization logic. The repaired pair is then rechecked and executed from a clean browser state. This process is repeated for at most \(K\) rounds and terminates once all tests pass; otherwise, the sample is discarded.

\textbf{Functional and visual verification.} Executing the repaired test script on the repaired webpage produces a structured test report \(R\) and a replayed state sequence
\begin{equation}
    \mathcal{S}^{(2)}
    =
    \left\{
        \hat{s}_0,\hat{s}_1,\ldots,\hat{s}_T
    \right\}.
\end{equation}
Functional correctness requires all specification actions and assertions to execute successfully. For a functionally valid pair, we further compare the replayed states with the recorded reference states:
\begin{equation}
    V
    =
    \frac{1}{T+1}
    \sum_{t=0}^{T}
    \operatorname{Sim}
    \left(
        s_t,\hat{s}_t
    \right),
\end{equation}
where \(\operatorname{Sim}(\cdot,\cdot)\) denotes the visual similarity function. If all tests pass but \(V<\delta\), the interaction semantics have already been validated by the test script; we therefore attribute the remaining visual mismatch to the webpage code and repair it using the mismatched screenshot pairs as evidence.

A synthesized sample is retained only if, within the repair budget, all test cases pass and the average visual similarity between the recorded and replayed states exceeds \(90\%\):
\begin{equation}
    \operatorname{Keep}
    =
    \mathbb{I}
    \left[
        \exists r\leq K:
        R^{(r)}=\mathrm{all\text{-}pass}
        \land
        V^{(r)}\geq 0.90
    \right].
\end{equation}
The resulting verified sample contains the task instruction, recorded interaction prototype, test specification, repaired webpage code, and executable Pytest script.


\textbf{Design Objective.} TEDS does not aim to recover the original website implementation or reproduce every reference state exactly. Instead, it constructs executable code-test pairs that remain mutually consistent during replay. The test script may adapt implementation-specific details, such as selectors, component structures, and synchronization, while the recorded states provide behavioral and visual guidance rather than exact oracles. A pair is valid when its tests execute successfully on the replayed states and the minimum visual-similarity requirement is satisfied; exact pixel-level, DOM-level, or implementation-level equivalence is not required.

Among the 2,058 collected tasks, 1,736 samples satisfy both the functional and visual requirements, resulting in an overall synthesis success rate of 84.35\%. 

\section{Test-Driven Post Training}

\subsection{Test ID Weighted Supervised Fine-Tuning}




Before reinforcement learning, we perform supervised fine-tuning (SFT) on the verified samples produced by TEDS. Each sample uses the recorded multi-state prototype as input, with the corresponding execution-verified webpage code as the target. A test ID is a unique HTML attribute assigned to an interactive element, providing a stable selector for test execution. For example, the button \texttt{<button data-testid="follow-button-salesforce">} can be located by the test script through \texttt{follow-button-salesforce}, regardless of its visual position or surrounding DOM structure. Test IDs therefore establish explicit bindings between test actions and generated webpage elements, enabling reliable interaction execution and reward computation. However, test-ID tokens constitute only a small fraction (less than 1\%) of the webpage code. Under standard token-level cross-entropy, their contribution can be overwhelmed by common HTML, CSS, and JavaScript tokens, causing the model to omit or incorrectly generate them. To address this imbalance, we apply a test-ID-weighted language modeling objective.  Specifically, we assign each token \(y_i\) a weight \(w_i=\lambda\) if \(y_i\in\mathcal{T}_{\mathrm{id}}\), and \(w_i=1\) otherwise, where \(\mathcal{T}_{\mathrm{id}}\) denotes the set of test-ID tokens and \(\lambda>1\) controls their importance. The weighted SFT objective is:

\begin{equation}
\mathcal{L}_{\mathrm{SFT}}
=
-\frac{1}{\sum_{i=1}^{N} w_i}
\sum_{i=1}^{N}
w_i
\log p_{\theta}
\left(
y_i \mid \mathcal{S},y_{<i}
\right).
\end{equation}
This objective preserves general webpage generation capability while explicitly encouraging complete and accurate test-ID generation, providing reliable element bindings for subsequent test-driven reinforcement learning.

\subsection{Test-Driven Group Relative Policy Optimization}

After SFT, we further optimize the model using Group Relative Policy
Optimization (GRPO). Given an interactive prototype \(\mathcal{S}\), the
policy samples \(G\) candidate webpages
\(\{\hat{W}_i\}_{i=1}^{G}\). Each candidate is rendered and evaluated using
its corresponding executable test script, avoiding costly GUI-agent
feedback. We construct the reward from two complementary signals.  (1) \textbf{Test pass rate.} We execute the test script against the generated webpage in a controlled browser environment. Given \(N\) functional checks, the reward is defined as  \(r_{\mathrm{pass}}=\frac{1}{N}\sum_{j=1}^{N}
\mathbb{I}[\mathrm{test}_j=\mathrm{pass}]\), measuring the proportion of successfully satisfied checks. (2) \textbf{Visual similarity.} We replay the interaction trajectory to
obtain \(\hat{\mathcal{S}}=\{\hat{s}_0,\ldots,\hat{s}_T\}\). The visual reward is defined as \(r_{\mathrm{vis}}=\frac{1}{T+1}\sum_{t=0}^{T} \operatorname{CLIPSim}(s_t,\hat{s}_t)\), measuring the average similarity between rendered and reference states. The final reward is \(r=\alpha_{\mathrm{pass}}r_{\mathrm{pass}}+
\alpha_{\mathrm{vis}}r_{\mathrm{vis}}\), where
\(\alpha_{\mathrm{pass}}\) and \(\alpha_{\mathrm{vis}}\) control their relative contributions.

 For each sampled group, GRPO normalizes the rewards
to obtain the relative advantage:
\begin{equation}
A_i
=
\frac{
r_i-\operatorname{mean}(\{r_j\}_{j=1}^{G})
}{
\operatorname{std}(\{r_j\}_{j=1}^{G})+\epsilon
}.
\end{equation}
Unlike actor--critic methods, this group-relative estimation does not require
an additional value model. For the \(t\)-th token of candidate
\(\hat{W}_i\), the probability ratio between the current and rollout
policies is defined as:
\begin{equation}
\rho_{i,t}(\theta)
=
\frac{
\pi_{\theta}
(\hat{w}_{i,t}\mid\mathcal{S},\hat{w}_{i,<t})
}{
\pi_{\mathrm{old}}
(\hat{w}_{i,t}\mid\mathcal{S},\hat{w}_{i,<t})
}.
\end{equation}

The policy is optimized using the clipped GRPO objective:
\begin{equation}
\resizebox{0.9\columnwidth}{!}{$\displaystyle
\mathcal{J}_{\mathrm{GRPO}}(\theta)
=
\mathbb{E}\!\left[
\frac{1}{G}\sum_{i=1}^{G}\frac{1}{L_i}\sum_{t=1}^{L_i}
\min\!\left(
\rho_{i,t}(\theta)A_i,
\operatorname{clip}\!\left(
\rho_{i,t}(\theta),1-\varepsilon,1+\varepsilon
\right)A_i
\right)
-
\beta D_{\mathrm{KL}}
\!\left(\pi_{\theta}\Vert\pi_{\mathrm{ref}}\right)
\right]
$}.
\end{equation}
Here, \(L_i\) is the generated code length, while \(\pi_{\mathrm{old}}\) and \(\pi_{\mathrm{ref}}\) denote the rollout and frozen SFT policies. The coefficients \(\varepsilon\) and \(\beta\) constrain policy updates and deviation from the reference, respectively. This objective jointly improves functional correctness and visual fidelity.


\section{Experimental Setup}

\subsection{Research Questions}

\begin{itemize}[leftmargin=*]
    \item \textbf{RQ1: Data Synthesis Performance}. How effectively does the proposed data synthesis pipeline synthesize clean and verified training data?
    \item \textbf{RQ2: Effectiveness}. Does \scheme \ outperform existing baselines in interactive webpage generation?

    \item \textbf{RQ3: Efficiency}. How much does \scheme \ reduce training costs?
    
    \item \textbf{RQ4: Parameter study}. How do key parameters affect the performance of \scheme?
   \item \textbf{RQ5: Case Study.} Why does \scheme \ work? 
\end{itemize}



\subsection{Evaluation Datasets}

We evaluate \scheme{} on two interactive webpage-generation benchmarks. (1) IWR-Bench~\cite{chen2026iwr} contains 113 tasks from 100 real-world websites, covering 1,001 actions. Each task provides a user-interaction video and static assets; we convert the video into interaction states and evaluate both functional correctness and visual fidelity. (2) Vision2Web~\cite{he2026visionweb} contains 193 tasks across 16 categories, with 918 prototype images and 1,255 functional tests. We use its interactive frontend subset and report the corresponding visual and functional scores.

\subsection{Evaluation Metrics}


For each benchmark, we follow its official evaluation protocol.

\subsubsection{Functional Correctness.} \textbf{Interactive
Functionality Score (IFS)}. For IWR-Bench~\cite{chen2026iwr}, a Browser-Use-based executor replays the ground-truth interaction trajectory, with GPT-4o~\cite{hurst2024gpt4o} used to ground natural-language action descriptions to webpage elements. An action is successful only if it executes without operational errors and all associated assertions are satisfied. The assertions are evaluated by Gemini-2.5-Pro~\cite{comanici2025gemini2.5} using screenshots captured before and after the action. For a task containing \(N_{\mathrm{total}}\) actions, the \textbf{IFS} is $\mathrm{IFS}=\frac{N_{\mathrm{succ}}}{N_{\mathrm{total}}},$ where \(N_{\mathrm{succ}}\) denotes the number of successfully completed actions. The final IFS is macro-averaged across all tasks.


\textbf{Functional Score (FS)}.
Vision2Web~\cite{he2026visionweb} evaluates functionality through workflow
nodes containing test objectives, guided actions, and validation criteria.
A GLM-4.6V-based GUI verifier~\cite{vteam2025glm45v} executes these nodes,
with the score defined as
\(\mathrm{FS}=\frac{1}{N_f}\sum_{i=1}^{N_f}\mathbb{I}(f_i=\mathrm{pass})\).

\subsubsection{Visual Fidelity.}
IWR-Bench evaluates visual fidelity at successfully reached checkpoints
marked for visual evaluation. (1) \textbf{Low-level Fidelity Score (LFS).} At checkpoint \(i\), LFS averages OCR-based Levenshtein similarity~\cite{cui2025paddleocr} and DINOv2-based~\cite{oquab2023dinov2}
feature similarity: $S_{\mathrm{LFS},i}=\frac{S_{\mathrm{OCR},i}+S_{\mathrm{DINO},i}}{2}.$ (2) The \textbf{High-level Fidelity
Score (HFS)} is assigned by Gemini-2.5-Pro through holistic comparison of
the generated and reference screenshots. (3) The \textbf{Visual Fidelity Score (VFS)} is then computed as
$\mathrm{VFS}=\frac{1}{|\mathcal{I}_{v,\mathrm{succ}}|}\sum_{i\in\mathcal{I}_{v,\mathrm{succ}}}\left(
        0.5S_{\mathrm{LFS},i}
        +
        0.5S_{\mathrm{HFS},i}
    \right),$
where \(\mathcal{I}_{v,\mathrm{succ}}\) contains the successfully reached
visual checkpoints. (4) \textbf{Final Score.} IWR-Bench combines functional and visual quality as $\mathrm{Final}=0.7\,\mathrm{IFS}+0.3\,\mathrm{VFS}.$ 

For Vision2Web, the visual judge is instantiated with Gemini-3-Pro-Preview. It compares each rendered page with its reference prototype at the component level, considering layout, content, typography, color, and overall appearance. (1) \textbf{Visual Score (VS)}. Given \(N_v\) prototypes, its Visual Score (VS) is $\mathrm{VS}=\frac{1}{N_v}\sum_{j=1}^{N_v} \operatorname{Fidelity}(P_j,\hat{P}_j),$ where \(P_j\) and \(\hat{P}_j\) denote the reference and rendered prototypes, respectively. (2) \textbf{Final Score.} The benchmark reports the overall score as $\mathrm{Final}=\frac{\mathrm{VS}+\mathrm{FS}}{2}.$



\subsubsection{Efficiency.}

We evaluate efficiency using the per-sample time consumed by the three main stages of RL training: \emph{reward computation}, which evaluates generated webpages and produces feedback; \emph{policy update}, which performs model optimization; and \emph{rollout generation}, which samples candidate webpages. In addition, we measure the number of external MLLM tokens consumed by GUI-agent or VLM-based verification. Lower values indicate greater efficiency.



\subsection{Baselines}

We compare our method with three groups of baselines. First, we include WebVIA with Qwen3.5-9B~\cite{qwen35} and Qwen3.5-27B~\cite{qwen35} as agent-harness baselines, representing existing agentic workflows for webpage generation. Second, we evaluate open-source MLLMs at different model scales. The lightweight group includes Qwen3.5-9B~\cite{qwen35}, GLM-4.6V-Flash~\cite{vteam2025glm45v}, and Qwen3-VL-8B-Instruct~\cite{qwen3technicalreport}, while the larger-model group includes Qwen3.8-27B~\cite{qwen38}, Qwen3.5-27B~\cite{qwen35}, and Qwen3-VL-30B-A3B-Instruct~\cite{qwen3technicalreport}. Finally, we compare against strong closed-source models, including GPT-5.6-Luna~\cite{openai2026gpt56luna}, Claude-Sonnet-5~\cite{anthropic2026sonnet5}, and Gemini-3.7-Flash~\cite{google2026gemini37flash}. We also include WebGen-Agent~\cite{lu2026webgen} and WebGen-R1~\cite{jiang2026webgen} as representative RL-based methods for interactive webpage generation. We instantiate all RL-based methods with the same Qwen3.5-9B backbone. For each method, we otherwise follow the best-performing hyperparameter configuration reported in its original paper.



\subsection{Implementation}

We use Qwen3.5-9B~\cite{qwen35} as the base MLLM and conduct SFT and GRPO with MS-SWIFT~\cite{zhao2024swift}. During training, the maximum completion length is 16,384 tokens, and the maximum input image size is 1,003,520 pixels. For SFT, we perform full-parameter fine-tuning on all 1,736 verified samples for five epochs. We use the test-ID-weighted objective with \(\lambda=2\), a global batch size of 64, and fused AdamW with a learning rate of \(1\times10^{-6}\), weight decay of 0.01, and maximum gradient norm of 1.0. The maximum sequence length is 32,768 tokens, with a cosine schedule and a warmup ratio of 0.05. For GRPO, we use 1,100 training samples and optimize for 500 steps, sampling 32 candidate webpages per input. We set the learning rate to
\(1\times10^{-5}\), sampling temperature to 0.8, and reward weights to
\((\alpha_{\mathrm{pass}},\alpha_{\mathrm{vis}})
=(\mathrm{0.5},\mathrm{0.5})\). We apply LoRA with rank 32,
\(\alpha=64\), and dropout 0.05. GRPO uses group-level reward normalization, a KL coefficient of \(\beta=0.04\), and AdamW with a cosine schedule, a warmup ratio of 0.03, and maximum gradient norm of 1.0. The maximum sequence length is 20,480 tokens. All training is conducted on eight NVIDIA H100 GPUs using BF16 precision.

\section{Experimental Results}

\subsection{Performance of Data Synthesis (RQ1)}

\textbf{Overall Performance and Ablation}. We randomly sample 249 tasks for the ablation study. In Table~\ref{tab:data-engine-ablation}, CC-TSG improves the pass rate from 15.66\% to 28.92\% and the CLIP score from 11.75\% to 23.18\%, demonstrating the benefit of incorporating code context into test-script generation. FAR yields substantially larger gains, reaching an 89.56\% pass rate and an 82.88\% CLIP score. Combining both components
achieves the best quality, with a 91.16\% pass rate and an 85.56\% CLIP score. CC-TSG also reduces the average token cost from \$0.0350 to \$0.0343 without FAR and from \$0.0677 to \$0.0598 with FAR. This reduction results
from using only interaction-relevant HTML fragments and improving first-pass test quality, which reduces the number of samples requiring repair. CC-TSG and FAR provide complementary improvements in synthesis quality and cost.


\begin{table*}[ht]
\centering
\caption{Ablation results of the two key components.
\textbf{CC-TSG} denotes Code Context-enhanced Test Script Generation, and
\textbf{FAR} denotes Failure-Attributed Repair. The best and second-best
results are highlighted in bold and underlined, respectively.}
\label{tab:data-engine-ablation}
\small
\setlength{\tabcolsep}{6pt}
\renewcommand{\arraystretch}{0.9}

\begin{tabular}{cccccc}
\toprule
\textbf{CC-TSG} &
\textbf{FAR} &
\textbf{Variant} &
\shortstack{\textbf{Pass Rate(\%)}} &
\textbf{CLIP(\%)} &
\shortstack{\textbf{Avg. Token Cost}\textbf{(\$/sample)}} \\
\midrule

\no  & \no  & \(C_0\) &
15.66 & 11.75 & \underline{0.0350} \\

\yes & \no  & \(C_1\) &
28.92 & 23.18 & \textbf{0.0343} \\

\no  & \yes & \(C_2\) &
\underline{89.56} & \underline{82.88} & 0.0677 \\

\yes & \yes & \(C_3\) &
\textbf{91.16} & \textbf{85.56} & 0.0598 \\

\bottomrule
\end{tabular}
\end{table*}

\begin{table*}[ht]
\centering
\caption{Distribution of failure types under different data-engine configurations.
\textbf{CC-TSG} denotes Code Context-enhanced Test Script Generation, and
\textbf{FAR} denotes Failure-Attributed Repair.}
\label{tab:failure-type-distribution}
\small
\setlength{\tabcolsep}{7pt}
\renewcommand{\arraystretch}{0.9}

\begin{tabular}{@{}lrrrr@{}}
\toprule
\textbf{Failure Type} &
\textbf{Vanilla} &
\textbf{CC-TSG} &
\textbf{FAR} &
\textbf{CC-TSG + FAR} \\
\midrule

\texttt{TARGET\_DEFINITION\_ERROR}       & 50 & 41 & 14 & 13 \\
\texttt{TARGET\_REFERENCE\_ERROR}        &  1 &  1 &  0 &  0 \\
\texttt{SCRIPT\_DOM\_MISMATCH}           & 40 & 13 &  2 &  2 \\
\texttt{SCRIPT\_SPECIFICATION\_MISMATCH}      &  3 &  8 &  4 &  2 \\
\texttt{TARGET\_AVAILABILITY\_ERROR}     & 17 & 22 &  8 &  9 \\
\texttt{BEHAVIORAL\_MISMATCH}            & 99 & 92 &  1 &  0 \\
\midrule
\textbf{Total Failures}                  &
\textbf{210} & \textbf{177} & \textbf{29} & \textbf{26} \\
\bottomrule
\end{tabular}
\end{table*}

\textbf{Failure Type Analysis}. Table~\ref{tab:failure-type-distribution} presents a detailed breakdown of the failures produced by different data-engine configurations. Introducing CC-TSG reduces the total number of failures from 210 to 177. The largest improvement occurs for \texttt{SCRIPT\_DOM\_MISMATCH}, which decreases from 40 to 13, indicating that access to relevant webpage-code context helps the model generate selectors and actions that are more compatible with the underlying DOM. However, CC-TSG alone cannot correct errors in the generated webpage or resolve ambiguous behavioral failures, which remain the dominant failure type.

FAR produces a substantially larger improvement, reducing total failures from 210 to 29. In particular, \texttt{BEHAVIORAL\_MISMATCH} decreases from 99 to 1, demonstrating that failure attribution and targeted repair effectively resolve ambiguous errors in webpage behavior and test semantics. Combining CC-TSG and FAR further reduces the total to 26 failures, corresponding to an 87.6\% reduction over the vanilla setting. Most remaining failures are target-definition or target-availability errors, suggesting that the residual difficulty mainly lies in generating webpage elements that are correctly defined and interactable at runtime. Overall, the two components address complementary sources of error and jointly improve the reliability of synthesized training samples.


\begin{figure}[ht]
    \centering
    \begin{minipage}{0.48\linewidth}
        \centering
        \subfigure[Pass Rate.]{
            \includegraphics[width=0.45\linewidth]{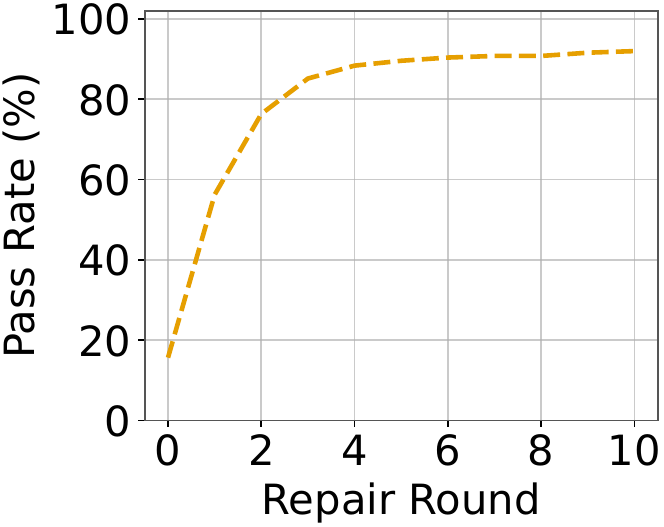}
        }
        \subfigure[CLIP Score.]{
            \includegraphics[width=0.45\linewidth]{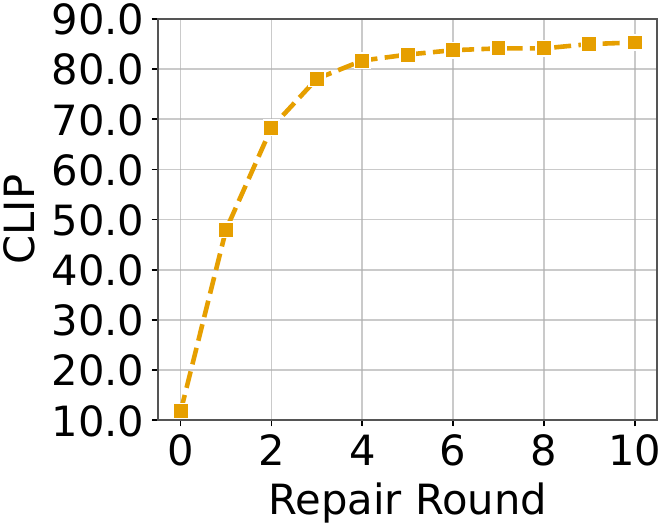}
        }
        \caption{Effect of Repair Rounds w/o Code Context.}
        \label{fig:round_c2}
    \end{minipage}
    \begin{minipage}{0.48\linewidth}
        \centering
        \subfigure[Pass Rate.]{
            \includegraphics[width=0.45\linewidth]{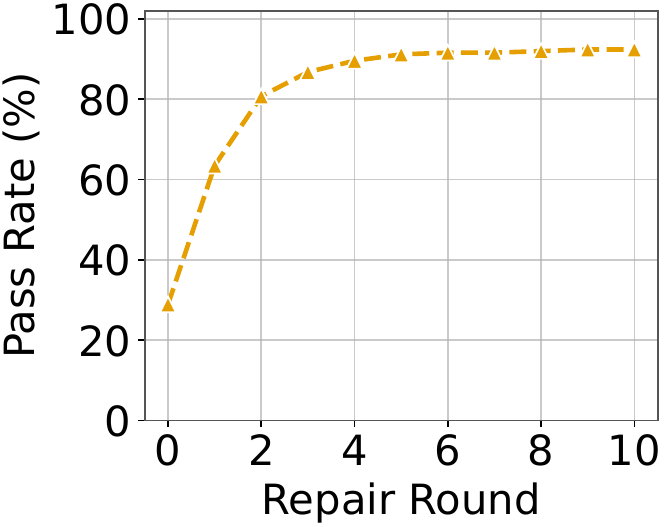}
        }
        \subfigure[CLIP Score.]{
            \includegraphics[width=0.45\linewidth]{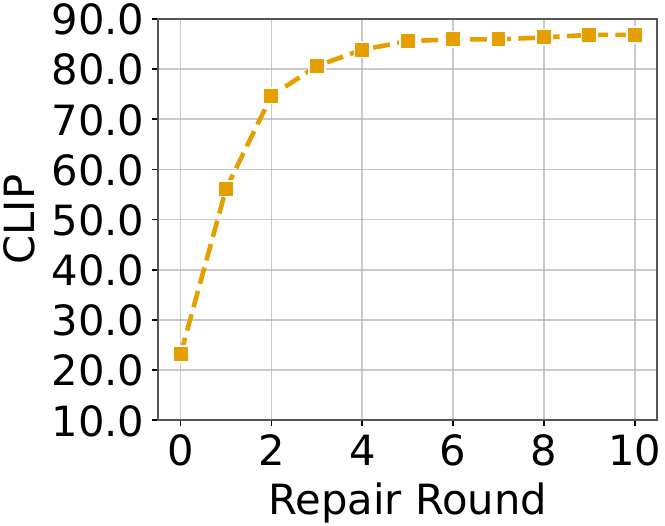}
        }
        \caption{Effect of Repair Rounds w/ Code Context.}
        \label{fig:round_c3}
    \end{minipage}
\end{figure}

\textbf{Repair Rounds Analysis}. Figures~\ref{fig:round_c2} and~\ref{fig:round_c3} analyze the effect of the maximum number of repair rounds. Iterative repair substantially improves both functional correctness and visual fidelity, with code context consistently yielding better initial and final performance. Most gains occur within the first four rounds, after which the curves begin to saturate and additional improvements become marginal. Therefore, we set the maximum number of repair rounds to \(K=4\), achieving a favorable trade-off between sample quality and generation cost.

\begin{tcolorbox}[
    colback=gray!20,
    colframe=gray!20,
    width=\columnwidth,
    left=0.05in,
    right=0.05in,
    top=0.05in,
    bottom=0.05in
]
\textbf{Answer to RQ1:}
CC-TSG and FAR jointly enable reliable data synthesis, achieving a 91.16\% pass rate, and 0.8556  CLIP score while reducing failures by 87.6\%. Most repair gains occur within four rounds.
\end{tcolorbox}

\begin{table*}[h]
\centering
\caption{Comparison with baseline methods on IWR-Bench and Vision2Web.
Higher values are better.}
\label{tab:main-results}
\small
\setlength{\tabcolsep}{1.5pt}
\renewcommand{\arraystretch}{0.9}

\begin{tabularx}{\textwidth}{
    @{}
    >{\raggedright\arraybackslash}p{3.6cm}
    *{8}{>{\centering\arraybackslash}X}
    @{}
}
\toprule
& \multicolumn{5}{c}{\textbf{IWR-Bench}}
& \multicolumn{3}{c}{\textbf{Vision2Web}} \\
\cmidrule(lr){2-6}
\cmidrule(lr){7-9}

\textbf{Method}
& \textbf{IFS}
& \textbf{LFS}
& \textbf{HFS}
& \textbf{VFS}
& \textbf{Final}
& \textbf{VS}
& \textbf{FS}
& \textbf{Final} \\
\midrule

\rowcolor{gray!12}
\multicolumn{9}{@{}c}{\textbf{Closed-Source Models}} \\

GPT-5.6-Luna
& 80.00 & 65.82 & 65.20 & 65.51 & 75.65
& 38.30 & 76.04 & 57.17 \\

Claude-Sonnet-5
& 44.83 & 65.02 & 64.99 & 65.00 & 50.88
& 36.17 & 58.99 & 47.58 \\

Gemini-3.7-Flash
& 38.48 & 62.69 & 63.38 & 63.03 & 45.85
& 50.01 & 36.02 & 43.01 \\

\addlinespace[2pt]
\rowcolor{gray!12}
\multicolumn{9}{@{}c}{\textbf{Open-Source Models: 30B Level}} \\

Qwen3.8-27B
& 42.19 & 68.82 & 69.57 & 69.19 & 50.29
& 30.27 & 42.31 & 36.29 \\

WebVIA (Qwen3.5-27B)
& 33.02 & 60.51 & 55.48 & 57.99 & 40.51
& 34.45 & 60.25 & 47.35 \\

Qwen3.5-27B
& 30.65 & 62.66 & 56.07 & 59.37 & 39.26
& 33.09 & 58.33 & 45.71 \\

\shortstack[l]{Qwen3-VL-30B-A3B-Instruct}
& 25.31 & 50.82 & 36.89 & 43.85 & 30.87
& 28.97 & 41.24 & 35.11 \\

\addlinespace[2pt]
\rowcolor{gray!12}
\multicolumn{9}{@{}c}{\textbf{Open-Source Models: 9B Level}} \\

WebVIA (Qwen3.5-9B)
& 22.45 & 45.19 & 42.36 & 43.78 & 28.85
& 31.54 & 42.11 & 36.83 \\


Qwen3.5-9B
& 20.62 & 38.24 & 29.89 & 34.06 & 24.65
& 28.90 & 42.19 & 35.56 \\

\shortstack[l]{GLM-4.6V-Flash}
& 12.74 & 48.14 & 38.05 & 43.10 & 21.85
& 23.30 & 28.52 & 25.91 \\

Qwen3-VL-8B-Instruct
& 9.38 & 27.37 & 17.84 & 22.60 & 13.35
& 26.80 & 22.67 & 24.74 \\

\addlinespace[2pt]
\rowcolor{gray!12}
\multicolumn{9}{@{}c}{\textbf{Existing RL-Based Methods}} \\

WebGen-R1~\cite{jiang2026webgen}
& 16.66 & 26.51 & 28.12 & 22.31 & 18.36
& 14.12 & 35.28 & 24.70 \\

WebGen-Agent~\cite{lu2026webgen}
& 15.00 & 27.44 & 20.58 & 24.01 & 17.7
& 13.62 & 29.31 & 21.47 \\

\addlinespace[2pt]
\rowcolor{blue!10}
\multicolumn{9}{@{}c}{\textbf{Ours (Qwen3.5-9B as Backbone)}} \\

InterWebCoder(SFT)
& 15.77 & 41.82 & 35.39 & 38.60 & 22.62
& 32.46 & 37.17 & 34.82 \\

InterWebCoder(GRPO)
& 31.48 & 45.55 & 39.89 & 42.72 & 34.85
& 29.19 & 48.86 & 39.03 \\

\rowcolor{blue!5}
InterWebCoder(SFT+GRPO)
& 46.38 & 60.73 & 46.47 & 53.60 & 48.54
& 37.52 & 59.31 & 48.42 \\

\bottomrule
\end{tabularx}
\end{table*}

\subsection{Performance of Test-Driven Post Training (RQ2)}



\textbf{Comparison with Baselines}.
As shown in Table~\ref{tab:main-results}, InterWebCoder(SFT+GRPO) achieves final scores of 48.54 on IWR-Bench and 48.42 on Vision2Web, improving the Qwen3.5-9B backbone by 23.89 and 12.86 points, respectively. Despite using only a 9B backbone, it outperforms several 30B-level models and the closed-source Gemini-3.7-Flash. Its performance is also comparable to Claude-Sonnet-5, trailing by 2.34 points on IWR-Bench while leading by 0.84 points on Vision2Web. On Vision2Web, it achieves the highest final score among all evaluated open-source models.

\textbf{Effect of SFT and GRPO}. SFT improves visual fidelity but degrades functional performance on both benchmarks, indicating that token-level imitation cannot directly optimize interaction correctness. GRPO alone raises the final scores to 34.85 on IWR-Bench and 39.03 on Vision2Web by introducing executable functional feedback. Combining both stages further improves these scores to 48.54 and 48.42, exceeding GRPO alone by 13.69 and 9.39 points, respectively. These results show that SFT provides a suitable initialization for webpage and test-ID generation, while GRPO directly improves interactive behavior through functional and visual rewards.

\textbf{Analysis of RL-Based Baselines}. WebGen-R1 underperforms because its execution- and VLM-based rewards do not directly verify the fine-grained state transitions required by interactive webpage generation. WebGen-Agent relies on a complete iterative agent loop for code generation, execution, feedback, and refinement, and thus transfers poorly outside this workflow. Moreover, both depend on costly and potentially noisy external model-based verification, with GUI agents being particularly susceptible to grounding, execution, and state-interpretation errors. \scheme \ instead uses deterministic executable tests aligned with the target interaction trajectory.


\begin{figure}[h]
    \subfigure[Reward.]{
    \centering
    \includegraphics[width = .23\textwidth]{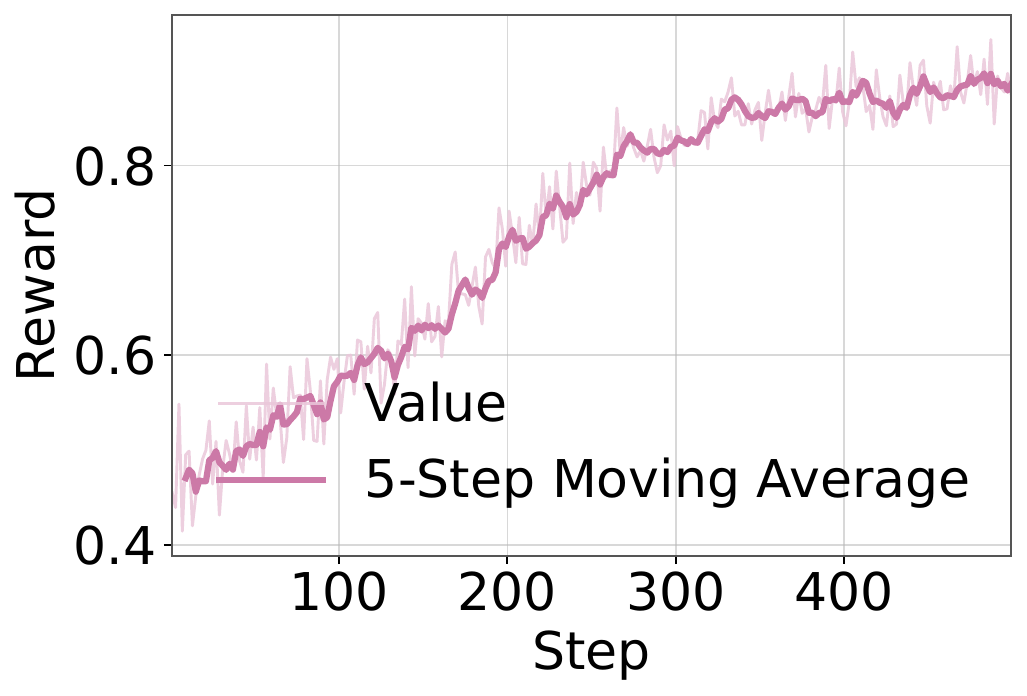}
    }    
    \subfigure[Policy Entropy.]{
    \centering
    \includegraphics[width = .23\textwidth]{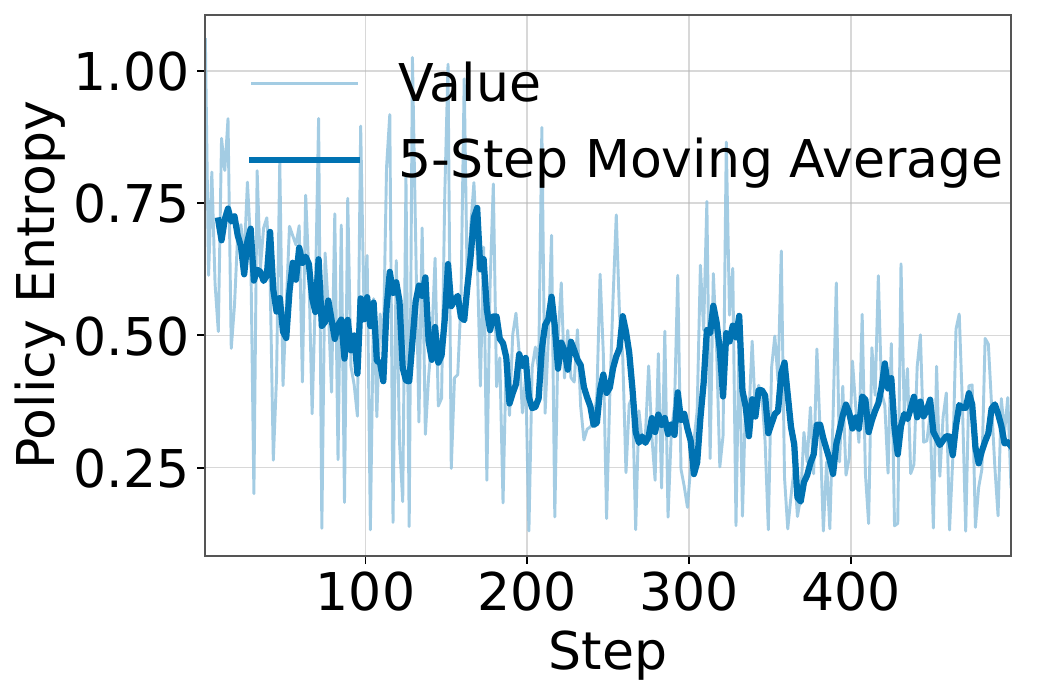}
    }
    \subfigure[KL Divergence.]{
    \centering
    \includegraphics[width = .23\textwidth]{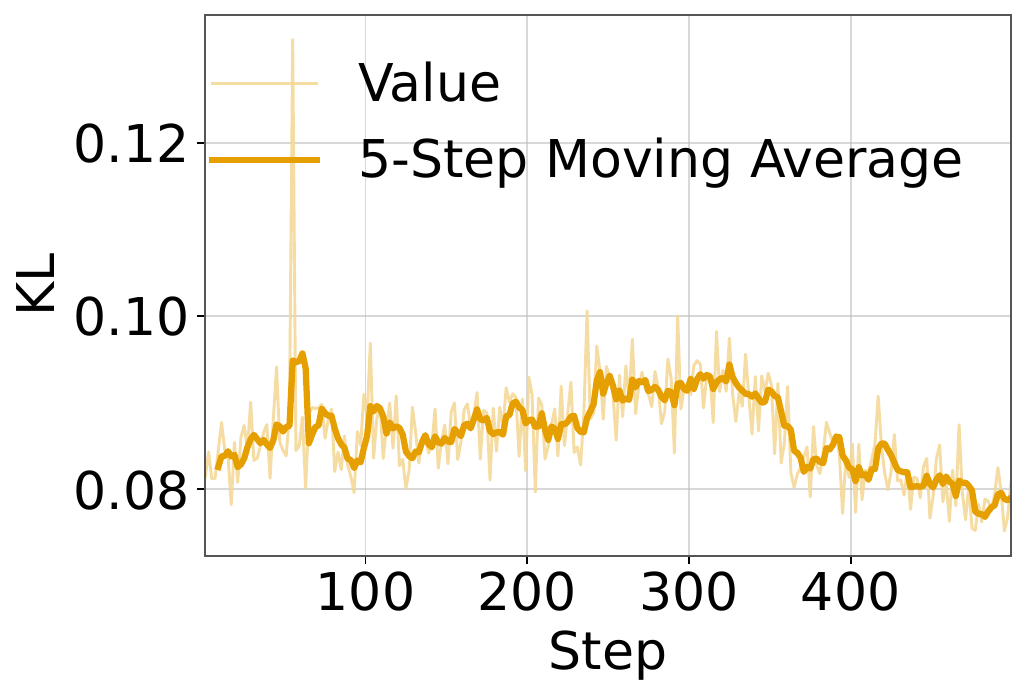}
    }
    \subfigure[Grad Norm.]{
    \centering
    \includegraphics[width = .23\textwidth]{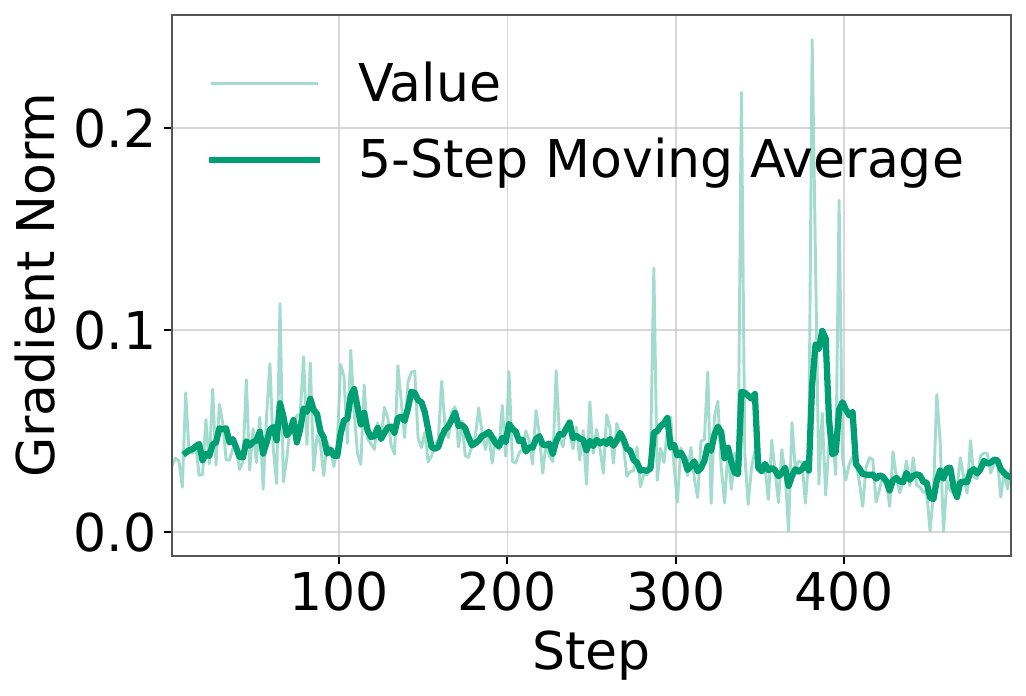}
    }
    \caption{Training dynamics of TDRL.}
    \label{fig:rq2_process}
\end{figure}

\textbf{GRPO Training Process Analysis.}
Figure~\ref{fig:rq2_process} presents the optimization dynamics of TDRL. The reward increases steadily from approximately 0.45 to 0.89 and gradually converges, indicating that the policy learns to generate webpages that better satisfy the executable functional and visual feedback. Meanwhile, policy entropy decreases from around 0.7 to 0.3, suggesting that the model becomes more confident in high-reward generation patterns while retaining some exploration. The KL divergence remains within a narrow range throughout training and declines toward the end, showing that the policy is updated without deviating excessively from the reference model. Finally, the gradient norm remains generally stable despite several transient spikes, with no evidence of gradient explosion. Overall, these trends demonstrate that test-driven rewards provide an effective and stable optimization signal for interactive webpage generation.


\begin{tcolorbox}[
    colback=gray!20,
    colframe=gray!20,
    width=\columnwidth,
    left=0.05in,
    right=0.05in,
    top=0.05in,
    bottom=0.05in
]
\textbf{Answer to RQ2:}
With a 9B backbone, \scheme{} surpasses several 30B-level and closed-source
models and achieves performance comparable to Claude-Sonnet-5.
\end{tcolorbox}

\subsection{Efficiency Study (RQ3)}

\begin{table*}[h]
\centering
\caption{
Efficiency comparison of different RL feedback mechanisms.
Time is reported in seconds per sample, with its percentage of the overall time in parentheses. Lower values are better.
}
\label{tab:feedback-efficiency}
\small
\setlength{\tabcolsep}{1.5pt}

\resizebox{\textwidth}{!}{
\begin{tabular}{
    @{}
    l
    l
    cccc
    c
    @{}
}
\toprule
\textbf{Method} &
\textbf{Feedback Mechanism} &
\textbf{Reward Time} &
\textbf{Update Time} &
\textbf{Rollout Time} &
\textbf{Overall Time} &
\textbf{Token Costs}\\
\midrule

WebGen-Agent GRPO~\cite{lu2026webgen} &
Screenshot + GUI agent &
130.01 (33.16\%) &
\underline{139.60 (35.61\%)} &
\textbf{122.40 (31.22\%)} &
392.01 &
16,699 \\

WebGen-R1 GRPO~\cite{jiang2026webgen} &
Execution + VLM judge &
92.40 (25.16\%) &
142.00 (38.66\%) &
\underline{132.90 (36.18\%)} &
367.30 &
\underline{6,684} \\



\rowcolor{blue!10}
\textbf{\scheme} &
Rule-based + Test script&
\textbf{2.08 (0.74\%)} &
\textbf{138.70 (49.33\%)} &
140.40 (49.93\%) &
\textbf{281.18} &
\textbf{0} \\

\bottomrule
\end{tabular}
}
\end{table*}

We compare \scheme{} with WebGen-Agent and WebGen-R1 under the same
Qwen3.5-9B setup. As shown in Table~\ref{tab:feedback-efficiency},
\scheme{} reduces reward computation from 130.01 and 92.40 seconds to 2.08 seconds per sample, a reduction of over 97\%. Because tests are executed locally, it also eliminates the external MLLM tokens required by WebGen-Agent (16,699) and WebGen-R1 (6,684) per sample. Consequently, the overall iteration time decreases to 281.18 seconds, yielding \(1.39\times\) and \(1.31\times\) speedups, respectively. These results demonstrate that executable tests remove the primary verification bottleneck in agent-based RL.

\begin{tcolorbox}[
    colback=gray!20,
    colframe=gray!20,
    width=\columnwidth,
    left=0.05in,
    right=0.05in,
    top=0.05in,
    bottom=0.05in
]
\textbf{Answer to RQ3:}
\scheme{} reduces reward latency by over 97\%, eliminates external MLLM costs, and achieves \(1.3\times\)-\(1.4\times\) end-to-end training speedups.
\end{tcolorbox}

\subsection{Parameter Study (RQ4)}


We study two key design choices in the SFT stage: the test-ID token loss weight \(\lambda\) and the interaction-step distribution of the training data. We report five metrics: \textbf{Code Integrity}, the proportion of complete and executable outputs; \textbf{Avg. ID Coverage}, the average proportion of required test IDs present; \textbf{Complete ID Coverage Rate}, the proportion of outputs containing all required test IDs; \textbf{Pass Rate}, the proportion of passed test steps; and \textbf{CLIP}~\cite{radford2021learning} for visual similarity.


\subsubsection{The Loss Weight \(\lambda\).}

\begin{figure}[h]
    \subfigure[Code Integrity.]{
    \centering
    \includegraphics[width = .23\textwidth]{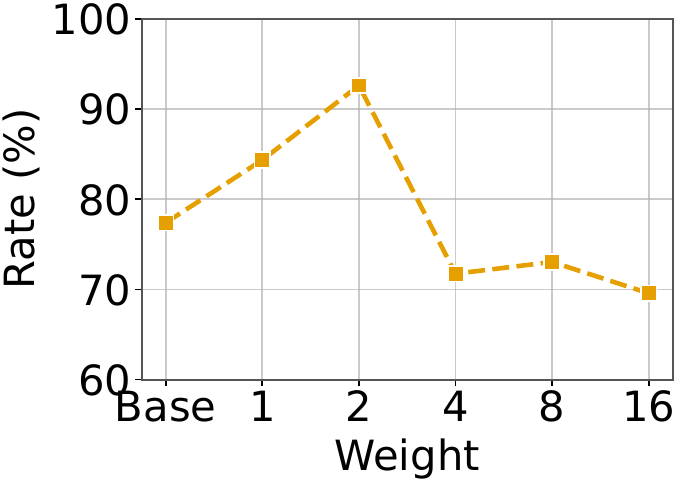}
    }    
    \subfigure[Test ID Coverage.]{
    \centering
    \includegraphics[width = .23\textwidth]{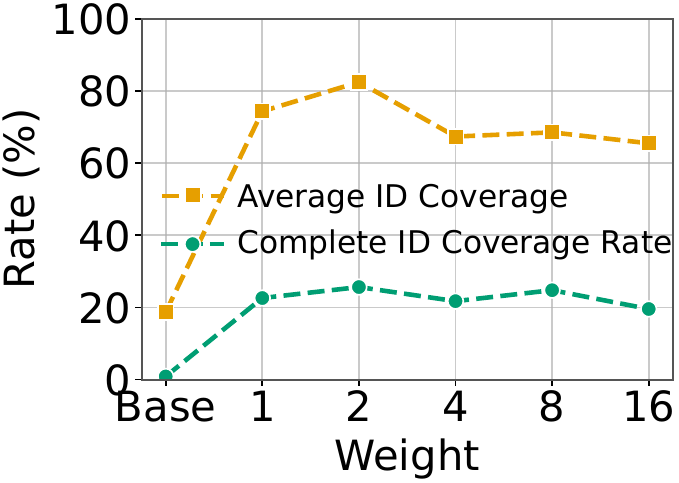}
    }
    \subfigure[Pass Rate.]{
    \centering
    \includegraphics[width = .23\textwidth]{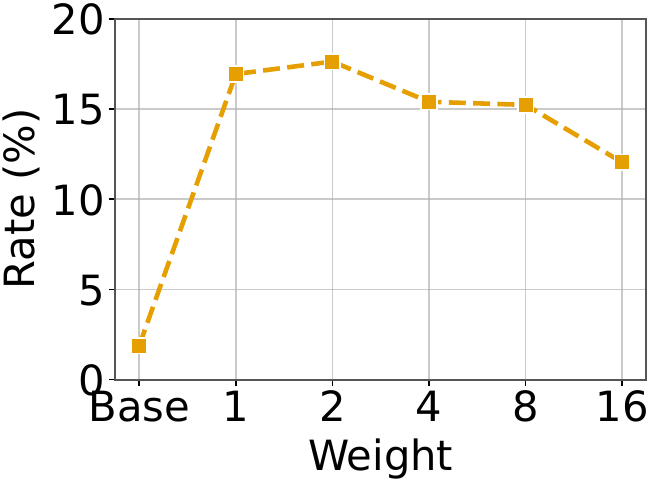}
    }
    \subfigure[Visual Similarity.]{
    \centering
    \includegraphics[width = .23\textwidth]{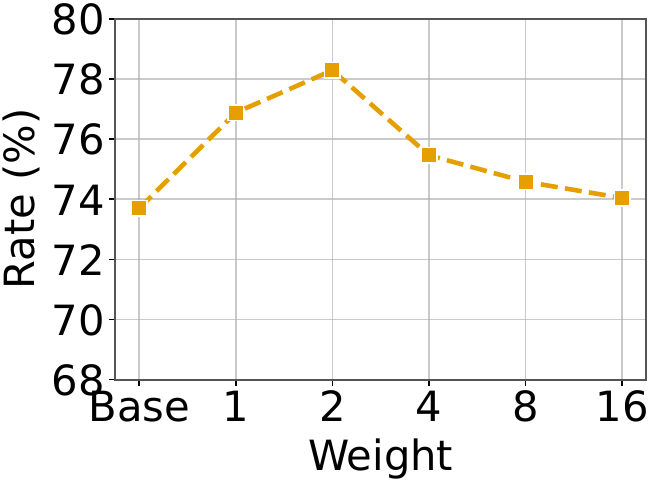}
    }
    \caption{Test token weight $\lambda$ analysis.}
    \label{fig:rq4_loss_weight}
\end{figure}

We randomly sample 230 examples from our synthesized dataset. Fig.~\ref{fig:rq4_loss_weight} examines the effect of the test-ID token weight \(\lambda\) on these samples. Increasing \(\lambda\) from 1 to 2 consistently improves code integrity, test-ID coverage, test pass rate, and CLIP similarity. In particular, \(\lambda=2\) achieves the highest metrics, demonstrating that moderately emphasizing test-ID tokens improves both code completeness and reliable test bindings. Further increasing \(\lambda\) generally degrades functional performance, as excessive weighting may overemphasize test IDs at the expense of other code tokens. We therefore set \(\lambda=2\), which provides the best overall balance between functional correctness and visual fidelity.


\subsubsection{The Interaction Step Range}

We study the effect of interaction length by dividing the SFT data into short (1--6), medium (7--10), long (11--14), and very long (15--46) trajectories, with 200 training examples per range. We also construct a 200-example mixed set with 50 examples from each range. All Qwen3.5-9B variants share the same setup and are evaluated on 80 held-out examples. As shown in Table~\ref{tab:interaction-step-range}, all ranges improve over the base model, but single-range training generalizes less effectively. The mixed setting achieves the highest average and complete ID coverage
(85.80\% and 22.50\%) and pass rate (14.10\%), while retaining competitive visual fidelity. We therefore adopt mixed-length trajectories.









\begin{table*}[h]
\centering
\caption{
Effect of the interaction-step range of the SFT data.
}
\vspace{-8pt}
\label{tab:interaction-step-range}
\small
\setlength{\tabcolsep}{2pt}
\renewcommand{\arraystretch}{0.9}
\begin{tabular}{lcccccc}
\toprule
\textbf{Step Range} &
\textbf{Code Integrity} &
\textbf{Avg. ID Coverage} &
\textbf{Complete ID Coverage Rate} &
\textbf{Pass Rate} &
\textbf{CLIP} \\
\midrule

Base (w/o SFT) &
\textbf{97.50} &
6.89 &
0.00 &
0.75 &
78.99 \\

Short (1--6) &
92.50 &
77.95 &
12.50 &
6.62 &
80.71 \\

Medium (7--10) &
88.75 &
\underline{84.19} &
16.25 &
7.59 &
\underline{81.98} \\

Long (11--14) &
86.25 &
79.01 &
12.50 &
6.94 &
81.69 \\

Very Long (15--46) &
87.50 &
80.64 &
\underline{18.75} &
\underline{10.47} &
\textbf{82.48} \\

Mix &
\underline{93.75} &
\textbf{85.80} &
\textbf{22.50} &
\textbf{14.10} &
81.90 \\

\bottomrule
\end{tabular}
\end{table*}

\begin{tcolorbox}[
    colback=gray!20,
    colframe=gray!20,
    width=\columnwidth,
    left=0.05in,
    right=0.05in,
    top=0.05in,
    bottom=0.05in
]
\textbf{Answer to RQ4:}
A moderate test-ID token weight of \(\lambda=2\) provides the best balance between functional correctness and visual fidelity. Moreover, mixed-length interaction trajectories generalize better than training data from any single step range.
\end{tcolorbox}

\subsection{Case Study (RQ5)}

\begin{figure}[h]
\centering
\includegraphics[width = .98\textwidth]{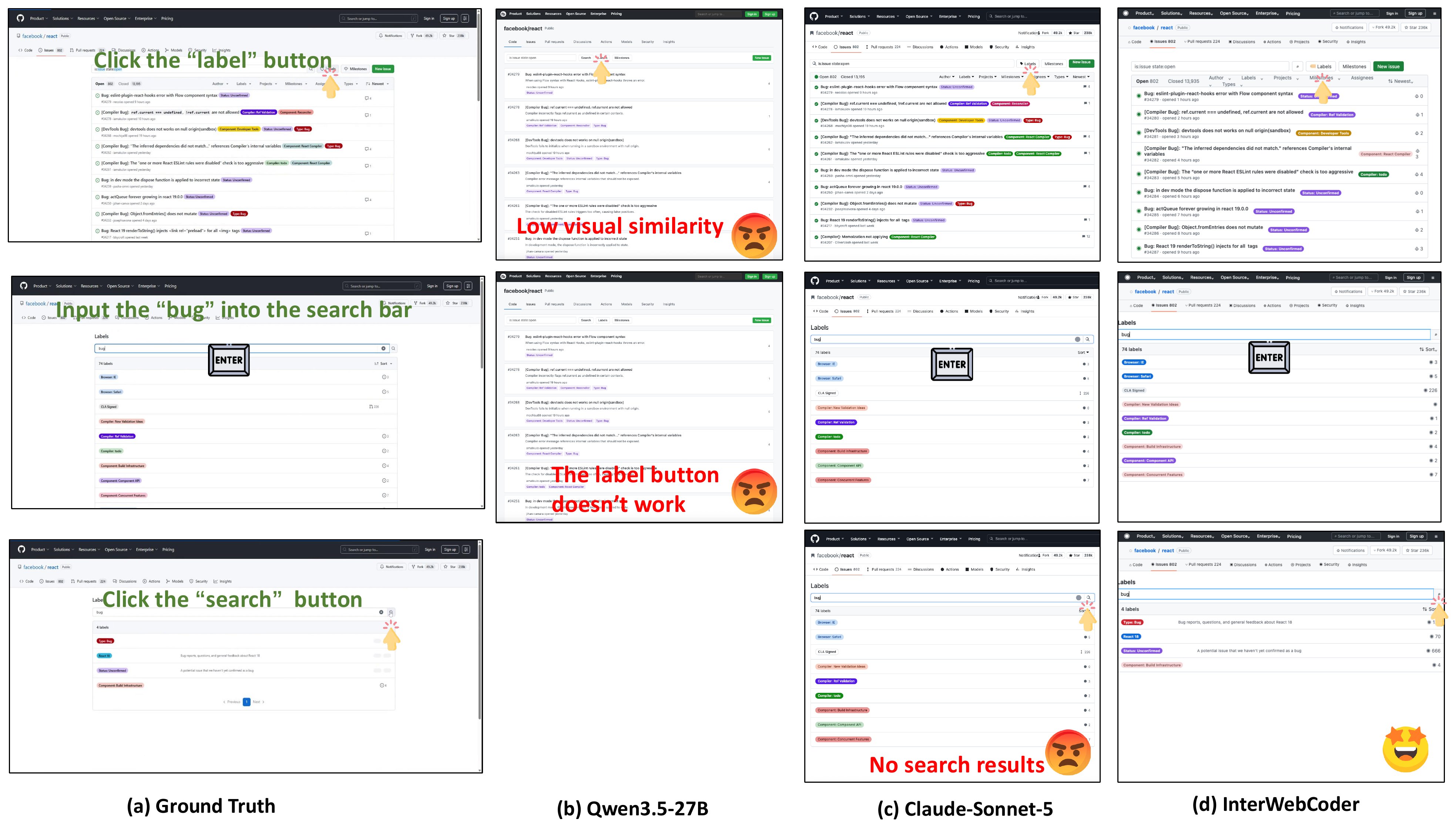}
\caption{Case study.}
\vspace{-10pt}
\label{fig:case}
\end{figure}

Figure~\ref{fig:case} presents a representative GitHub label-search task consisting of three actions: opening the label page, entering ``bug'' into the search bar, and executing the search. Qwen3.5-27B produces a webpage with low visual similarity, and its label button fails to trigger the required navigation.
Claude-Sonnet-5 more closely reconstructs the interface and supports text input, but the search action returns no results. In contrast, \scheme{} successfully reproduces all three state transitions and displays the expected filtered labels while maintaining high visual fidelity to the reference. This improvement stems from our GRPO reward design. The test-pass reward promotes correct navigation, input, and result filtering, while the visual reward aligns each rendered state with its reference. Together, they optimize both interaction behavior and visual fidelity.

\begin{tcolorbox}[
    colback=gray!20,
    colframe=gray!20,
    width=\columnwidth,
    left=0.05in,
    right=0.05in,
    top=0.05in,
    bottom=0.05in
]
\textbf{Answer to RQ5:}
GRPO jointly optimizes test execution and state-level visual similarity,
enabling \scheme{} to reproduce complete and visually faithful interactions.
\end{tcolorbox}

\section{Threats to Validity}


(1) \textit{Generalizability.} Our data cover 275 websites across 15 domains and include interaction trajectories of varying lengths and action types. Nevertheless, they cannot represent every web application. We reduce this threat by evaluating on both IWR-Bench and Vision2Web, which contain interaction scenarios and website categories beyond our synthesized data. (2) \textit{Backbone generality.} Due to computational constraints, we train InterWebCoder using Qwen3.5-9B as the primary backbone. However, Qwen3.5-9B is a representative and widely adopted open-source VLM with strong visual understanding and code-generation capabilities, making it a practical backbone for validating our method. Evaluating additional backbones remains future work.


\section{Related Work}

\textbf{UI Code Generation Methods}.
Recent MLLM-based methods improve webpage generation through decomposition, structured representations, and iterative refinement. DCGen~\cite{wan2024automatically} adopts divide-and-conquer generation, while DeclarUI~\cite{zhou2024bridging}, UICopilot~\cite{gui2025UICopilot}, LayoutCoder~\cite{wu2024mllm}, and LatCoder~\cite{gui2025latcoder} exploit segmentation, hierarchical generation, or layout-aware representations. ComUICoder~\cite{xiao2026comuicoder} promotes reusable components, DesignRepair~\cite{designrepair} iteratively repairs generated interfaces, and EfficientUICoder~\cite{xiao2025efficientuicoder} reduces redundant visual and code tokens. Recent work extends UI-to-Code to interactive settings: WebVIA~\cite{xu2025webvia} collects multi-state screenshots through agent exploration, Video2Code~\cite{xu2026video2code} recovers state transitions from UI videos, and TDDev~\cite{wan2026runnable}, WebGen-Agent~\cite{lu2026webgen}, and WebGen-R1~\cite{jiang2026webgen} introduce test- or feedback-driven generation. However, these methods either focus on static interfaces or depend on costly agent-based verification, without a scalable data-synthesis and training pipeline.

\textbf{UI Code Generation Benchmarks}.
Existing benchmarks cover increasingly realistic web-development settings. Web2Code~\cite{yun2024web2code} studies HTML synthesis using largely synthetic data, whereas Design2Code~\cite{si2025design2code} and WebCode2M~\cite{gui2025webcode2m} provide 484 and 20,000 real-world webpage samples, respectively. MRWeb~\cite{wan2024mrweb} and DesignBench~\cite{xiao2025designbench} further evaluate multi-page and framework-based development. For interactive generation, Interaction2Code~\cite{xiao2024interaction2code} covers 97 webpages and 213 interactions, WebGen-Bench~\cite{lu2026webgen} evaluates multi-file websites with executable tests, and IWR-Bench~\cite{chen2026iwr} assesses visual and functional reconstruction from interaction videos. Vision2Web~\cite{he2026visionweb} provides a broader hierarchy spanning static pages, interactive frontends, and full-stack applications. These benchmarks reflect the shift from static visual similarity toward executable and stateful evaluation.

\textbf{Post-training for Web Generation}.
Post-training methods can be divided into static and interactive settings.
For static UI generation, UICoder~\cite{wu2024uicoder} uses compiler and VLM
feedback for data filtering and preference alignment, while
UI2Code\(^{\mathrm{N}}\)~\cite{yang2026uicodetextn} introduces Relative
Visual Policy Optimization. Interactive generation additionally requires
behavioral supervision. WebVIA~\cite{xu2025webvia} fine-tunes VLMs on
multi-state screenshots and executable code; WebGen-Agent~\cite{lu2026webgen}
uses screenshot judgments and GUI-agent testing for Step-GRPO; and
WebGen-R1~\cite{jiang2026webgen} combines structural, execution-grounded,
and VLM-based rewards. In contrast, \scheme{} unifies data synthesis and
post-training for multi-state interactive prototypes, using
execution-verified tests as efficient feedback without repeated GUI-agent or
VLM verification.

\section{Conclusion}

We present \scheme{}, an efficient framework for interactive webpage code generation. Its Test-Specification-Guided and Execution-Verified Data Synthesis pipeline converts real-world interaction trajectories into clean webpage code and executable tests, producing 1,736 verified samples with an 84.35\% synthesis success rate. Test-ID-Weighted SFT learns reliable element bindings, while Test-Driven GRPO directly optimizes test pass rate and state-level visual similarity without GUI-agent verification. On IWR-Bench and Vision2Web, \scheme{} achieves final scores of 48.54 and 48.42, outperforming several 30B-level and closed-source baselines with only a 9B backbone. It further reduces reward-computation latency by over 97\%, eliminates external MLLM verification costs, and accelerates end-to-end training by up to \(1.4\times\).



\bibliographystyle{ACM-Reference-Format}
\bibliography{sample-base}

\end{document}